\documentclass{acm_proc_article-sp}
\usepackage{textcomp}
\usepackage{url}

\def\sharedaffiliation{%
\end{tabular}
\begin{tabular}{c}}
\makeatletter  
\renewcommand{\verbatim@font}{
  \ttfamily\footnotesize\itshape\catcode`\<=\active\catcode`\>=\active }
\makeatother 
\usepackage{float}
\floatstyle{ruled}
\newfloat{program}{h}{lop}
\floatname{program}{Code}

\begin{document}

\title{Quality assessment of the MPEG-4 scalable video CODEC}

\numberofauthors{3}
    \author{
      \alignauthor Florian Niedermeier\\
      \email{niederme@fim.uni-passau.de}
      \alignauthor Michael Niedermeier\\
      \email{niedermm@fim.uni-passau.de}
            \alignauthor Harald Kosch\\
      \email{harald.kosch@uni-passau.de}
      \sharedaffiliation
      \affaddr{Department of Distributed Information Systems}\\
      \affaddr{University of Passau (UoP)}\\
      \affaddr{Passau, Germany}
}

\date{\today}

\maketitle
\sloppy
\begin{abstract}

In this paper, the performance of the emerging MPEG-4 SVC CODEC is evaluated. In the first part,
a brief introduction on the subject of quality assessment and the development of the MPEG-4 SVC
CODEC is given. After that, the used test methodologies are described in detail, followed by an explanation of the actual test scenarios.
The main part of this work concentrates on the performance analysis of the MPEG-4 SVC CODEC - both objective and subjective.

\end{abstract}




\section{Introduction}
As both high visual quality and low bandwidth requirements are key features in the emerging mobile multimedia sector, MPEG and VCEG introduced a new extension to the MPEG-4 AVC standard - scalable video coding (SVC)\footnote{The SVC reference software has gone into Final Draft International Standard in the MPEG October meeting 2008.}. Its focus lies on supplying different client devices with video streams suited for their needs and capabilities. This is achieved by employing three different scalability modes: Spatial, temporal and SNR scalability. Because these new features are still in development and their impact on visual quality has not often been independently tested, this paper covers this subject.\\
The performance evaluation is done using both objective and subjective assessment methods. Each method has different advantages:\\
While subjective testing reflects the viewers impressions best, it has several downsides. It is much more time consuming and therefore also expensive. Also, very small differences in video quality cannot be reliably detected.\\
In contrast, objective analysis can effectively be run automated on computer systems. Therefore, it is much cheaper and its results are easily comparable. However, the objective metrics used to calculate the quality scores do never perfectly reflect user experience.\\
By using both subjective and objective test methodologies, the advantages of each assessment method can be used to its full potential. The comparison of the results also expresses the deviation of the objective scores from the viewers' subjective opinion.\\
Additionally to the evaluations covering the matter of visual quality, additional test runs are performed to check the encoding speed of the SVC CODEC, which is also an important feature, especially when looking at realtime encoding scenarios.\\
The assessment is divided into two separate parts: The first one is a MPEG-4 SVC stand-alone test, which throughly examines the impact of different encoding settings on the CODEC's performance. The second part of the testing consists of a competitive comparison of the MPEG-4 SVC reference CODEC, x264 (MPEG-4 AVC based) and Xvid (MPEG-4 ASP based), to analyze each CODEC's advantages and disadvantages in different usage scenarios. All tests are described in detail in section~\ref{sec:evaluations}.

\section{Related work}
Most of today's quality evaluations are run objectively, because of the previously mentioned high complexity and costs of subjective assessments. Still, some comparisons of subjective and objective assessment methods have been conducted, especially the CS MSU Graphics \& Media Lab Video Group ran several evaluations concerning CODEC competitions featuring various MPEG-4 ASP \& AVC implementations \cite{comparison2007} \cite{comparison2006}.\\
The emerging MPEG-4 SVC standard, however, has not been tested in such a manner. Although both objective \cite{bb15205} and subjective tests \cite{bb9878} have already been run separately, an analysis offering both test methodologies was yet outstanding.
The results of the subjective evaluation of the SVC reference CODEC \cite{bb9878} are limited to the quality change if temporal levels are reduced in exchange for higher video quality.\\
Besides that, the MPEG-4 SVC CODEC was also evaluated in an official ISO test \cite{ISO_VERI}, which did however not assess a broad range of quality-impacting parameters, but only tested a few basic features. Another problem concerning this evaluation is that it only focused on the comparison of MPEG-4 SVC and its direct predecessor MPEG-4 AVC. No CODEC implementing the still commonly used MPEG-4 ASP standard was rated in the comparison, nor were the performance impacts of the encoding parameters of SVC evaluated.\\
In this paper, a broader range of quality-affecting settings and scenarios is assessed, including both a SVC stand-alone test as well as a competitive comparison of different CODECs, to provide a large-scale overview of the current SVC CODEC's performance. In the SVC stand-alone test, special attention is paid to the influences the new scalable features of the SVC CODEC have on the visual quality.\\
Additionally, a comparative synthesis that comprises both subjective and objective test methods is conducted in this work.

\section{Used test methodologies}
\label{sec:metrics}
To provide comparable results, it is important for both objective and subjective assessments to be run under strictly specified conditions. This means for objective tests that the used metric, which calculates the difference between an impaired and an original image, and the encoding parameters are kept throughout the whole assessment.\\
Additionally to the facts stated for objective testing, subjective evaluations also need to have a fixed testing setup and environment, as various influences, like noise or sunlight can bias a users' opinion.\\
The test methodologies used in the evaluation are throughly described in the following.

\subsection{Objective metrics}
\subsubsection{PSNR}
The PSNR is the currently most widely used metric for quality evaluations of compression techniques. The result is given in the logarithmic unit decibel (dB). Even though this metric can be calculated for luminance as well as chrominance channels, it is common to just use Y-PSNR, meaning only the difference in luminance is evaluated. PSNR is calculated using the following equation: $PSNR = 20\cdot lg\frac{255}{\sqrt{MSE}}$ where $MSE = \frac{1}{mn}\sum \limits_{i=0}^{m-1}\sum \limits_{j=0}^{n-1}\| X(i, j) - Y(i, j)\| ^{2}$. When calculating the PSNR for a sequence of pictures, the MSE is calculated for the entire sequence and then inserted in the formula above, instead of calculating the PSNR for each frame and then calculating the mean \cite{msu_psnr}. The correlation of PSNR to subjective quality impression is discussed controversially: The results of the video quality experts group \cite{vqeg_results} come to the conclusion that PSNR correlation is on par with that of other metrics. In contrast, newer tests like \cite{comparison2006} claim that the correlation of PSNR is significantly lower than that of the SSIM metric \cite{ssim2004}. Still, PSNR is the standard metric used in most quality assessments and literature. To ensure comparability, this metric will be used in the following tests too.

\subsubsection{PSNR adaption for temporally scaled videos}
As shown in \cite{temp_metric}, normal PSNR calculation is not suitable for quality assessment of videos with temporal scalability. The calculated values are too low to accurately reflect perceived quality, so the following adapted quality score based on PSNR was proposed: $QM = PSNR + m^{0{.}38} \left(30 - FR\right)$. $QM$ is the metrics score, $FR$ is the framerate of the processed video. To calculate $PSNR$ in this equation, the frames of the temporally scaled video are repeated to match the frame count of the original sequence. The resulting sequence is then compared to the original using standard PSNR calculation. The parameter $m$ is the normalized average magnitude of large motion vectors, which is used to measure motion speed. The large motion vectors are the top 25\% of the largest motion vectors in the video sequence. After calculating the average magnitude of the large motion vectors, this value is normalized by the image width \cite{temp_metric}.
The equation was specifically designed for videos with a maximum framerate of 30 Hz. As the source videos used in the following work have different framerates, the following has to be considered: A simple adaption of the equation to fit the new source framerate ($QM = PSNR + m^{0{.}38} \left(60 - FR\right)$) does not lead to reasonable results, so the impact of temporal decimation is only considered if the framerate drops below 30 Hz. This means that sequences with a framerate of 30 Hz or lower are always compared against those with 30 Hz, so the metric described in \cite{temp_metric} can be used without modification.

\subsection{SAMVIQ}
The Subjective Assessment Methodology for VIdeo Quality (SAMVIQ) is an invention of the EBU (European Broadcasting Union), which started in 2001 and finished in 2004. It is incorporated in ITU-R BT.700   by now \cite{SAMVIQ-EBU}. SAMVIQ was developed because most other subjective test methodologies (for example DSIS, DSCQS, SSCQE and SDSCE) are specialized in rating videos shown on TV screens, and not on home computer or even mobile devices.\\
At the beginning of the test process, the subject watches the reference sequence. After that the expert has to watch and rate all impaired sequences, which are randomly ordered and made anonymous to the expert by labeling them alphabetically. If required, every sequence may be repeated as often as the tester likes. It is also possible to change the rating of a sequence anytime. The reference is also hidden among the impaired sequences and is therefore rated as well.\\
For voting, a linear, continuous scale with a range of 0 to 100 points is used, where a higher value represents better image quality and a lower one worse quality respectively \cite{SAMVIQ-EBU} \cite{INST-RUND}.\\

\begin{figure}[H]
\centering
\includegraphics[width=0.45\textwidth]{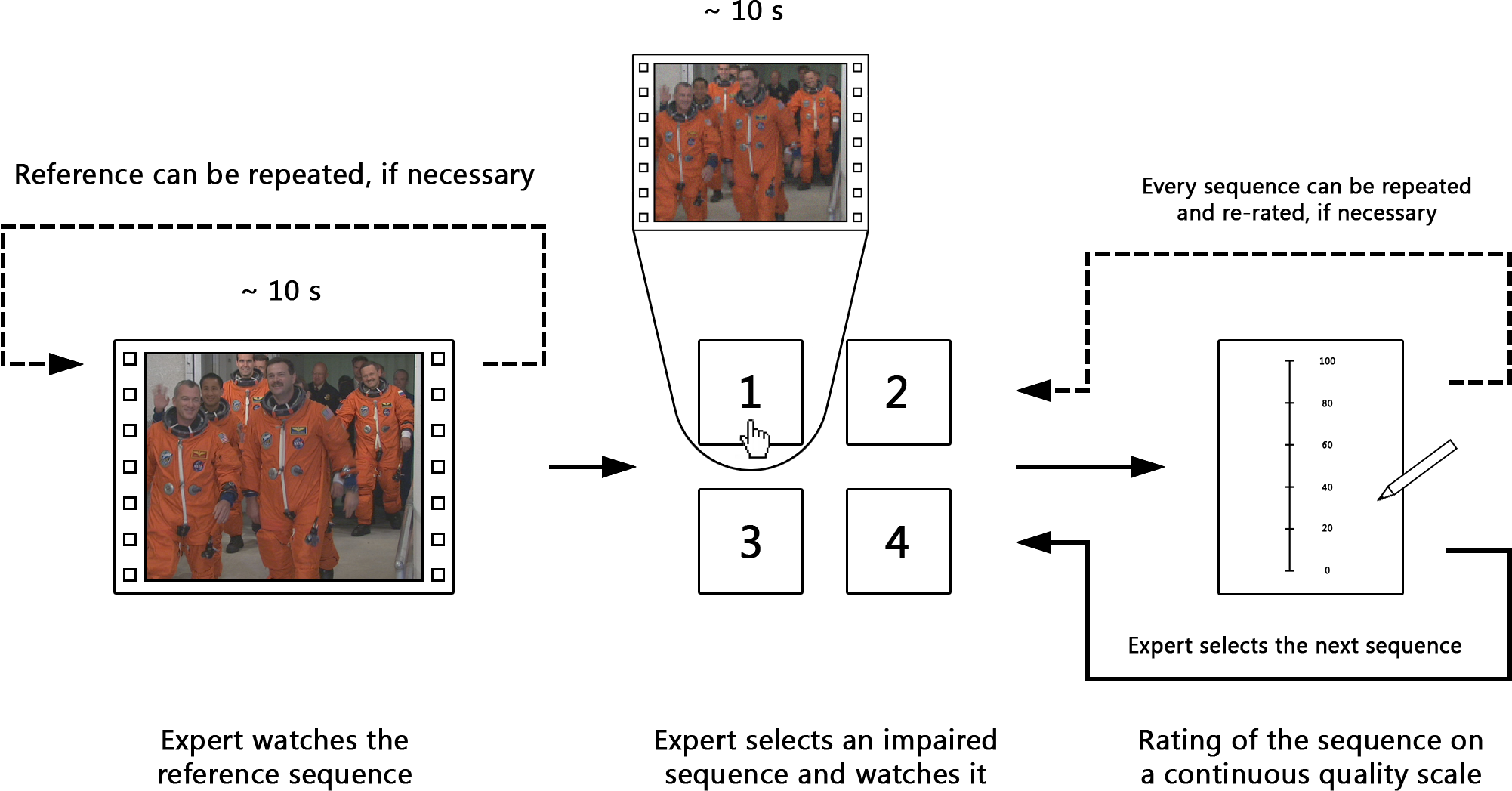}
\caption{Schematic of the SAMVIQ test methodology.}
\label{fig:SSCQE}
\end{figure}

\section{Test setup}
\label{sec:setup}
\subsection{Selection of experts}

The people that participate in the subjective assessment did not undergo a special selection. So a total of 21 persons of all age and working classes are included in the test. None of the experts was previously trained as a subjective tester or had a job associated with some kind of visual quality testing.\\
However, before a person is approved as an expert in the evaluation, two aptitude tests are run: A visual acuity and a color blindness test.\\
Visual acuity is obviously of great importance in subjective assessments of video quality, because even small differences in visual quality have to be detected by the expert. For that reason, the visual acuity of every viewer is tested using the Freiburg Visual Acuity, Contrast \& Vernier Test (FrACT). This free program allows a computer-aided check of i.a. the visual acuity while complying with the EN ISO 8596 standard. The process is thoroughly described in \cite{FRACT}. An acuity minimum of 1.0 is necessary, to take part in the following quality evaluation. Vision aids, like glasses or contacts, are permitted in the test.\\
The color perception is also an important factor when assessing graphical material. Persons with a visual impairment of the color perception (like red-green (protanopia, deuteranopia), blue-yellow (tritanopia) or total color blindness(achromatopsia)) cannot reliably detect color aberrations, which are a common error in video compression, and are therefore excluded from the test \cite{color}. This test is executed using the standard Ishihara test charts.\\
After these tests, one person had to be excluded, leaving a total of 20 test subjects for the subjective assessment.\\

\subsection{Subjective test environment}
The testing environment is set up as follows:\\
To prevent any unwanted display-related influences, the same device (a Samsung R40-T5500 Cinoso notebook, further technical details are shown in table~\ref{tab:test_sys_2} of section~\ref{speed}) is used for every test session and expert. The LCD supports a resolution up to 1280$\times$800 pixels and a luminance up to 200 cd/m$^2$. The black level and contrast of the display are adjusted using a PLUGE (Picture Line-Up Generation Equipment) pattern. PLUGE patterns vary in format, but a typical pattern consists of at least three vertical bars (called PLUGE pulses) with different shades of black and dark gray. The adjustment process and the generation of a PLUGE pattern itself is not described here in more detail, further information concerning this is provided in \cite{PLUGE}.\\
During the playback of the sequences the test room's background lighting is provided by a faint, artificial light source. The display is protected from direct light irradiation, to eliminate reflections. Daylight and other outside influences are also avoided as much as possible. The viewing distance is set concerning the rules of Preferred Viewing Distance (PVD) for an 15.4'' LCD device.\\
The display is aligned both horizontally and vertically to provide a viewing angle of $\le$~20\textdegree ~to the expert, which is well inside the recommended parameters (viewing angle $\le$~30\textdegree) stated in \cite{ITU-BT500}.

\subsection{Encoder settings}

Three CODECs are assessed in the comparison: Xvid 1.1.3 (MPEG-4 ASP), x264 core 59 r808bm ff5059a (MPEG-4 AVC) and the new MPEG-4 SVC reference encoder 9.12.2. All encoder parameters are kept at default settings except for the settings listed in tables \ref{tab:x264_settings} and \ref{tab:svc_settings}.

\begin{table}[H]
\centering
\begin{tabular}{|p{3.5cm}|p{3.5cm}|}
\hline
Encoding Type&Single pass - bitrate-based (ABR)\\
Max consecutive&2\\
Threads&4\\\hline
\end{tabular}
\caption{x264 encoder settings.}
\label{tab:x264_settings}
\end{table}

\begin{table}[H]
\centering
\begin{tabular}{|p{3.5cm}|p{3.5cm}|}
\hline
GOPSize&4\\
SearchMode&4\\
BaseLayerMode&2\\\hline
\end{tabular}
\caption{SVC encoder settings.}
\label{tab:svc_settings}
\end{table}

The 'GOPSize' parameter is changed to a value of 4 to enable the usage of B frames. Encoding a video sequence without B frames would result in a significant drop in compression efficiency.\\
The fast search algorithm is used, so 'SearchMode' is adjusted to 4.\\
The parameter 'BaseLayerMode' is altered as the default setting is invalid.

\section{Conducted evaluations}
\label{sec:evaluations}
The assessment is split in two separate evaluations: Firstly, the MPEG-4 SVC CODEC is tested in a stand-alone test, to document the impact of different encoder settings on the resulting quality and assess the CODEC's features.\\
Secondly, the characteristics of the MPEG-4 SVC CODEC are compared to those of x264 and Xvid in a comparison test.

\subsection{MPEG-4 SVC stand-alone test}

\subsubsection{Encoding parameter test}
\paragraph{Comparison of different block matching metrics}
First, the different metrics available for block matching will be compared. There are four different options available for FullPel and SubPel:
\begin{itemize}
    \item{SAD}
    \item{SAD-YUV (Not available for SubPel estimation)}
    \item{SSE}

    \item{HADAMARD}
%
\end{itemize}
Metrics can be chosen independently for FullPel and SubPel calculations. In this test, the different metrics are compared in terms of impact on encoding speed and visual quality.

\paragraph{Effect of different block matching algorithms and search parameters}
There are two different options for block matching algorithms available: Block search and fast search. Block search is an algorithm usually referred to as full search or exhaustive search. Without restraints in search range, it offers perfect prediction, but at the cost of extremely high computational complexity, as all possible blocks have to be compared. The second option named fast search is - as the name indicates - a much faster alternative. The developers claim that the loss of precision is by far outweighed by the speed increase obtained by this algorithm \cite{soft_manual}. This test will show how the combinations of search range parameters, defined by the variables 'SearchRange', 'BiPredIter' and 'IterSearchRange' in the encoder configuration, and block matching algorithms perform in terms of encoding speed and visual quality.

\subsubsection{Quantization parameter test}During this test, the impact of the quantization parameter (QP) on the video quality is evaluated. The value of the QP changes the strength of the quantization: The higher the QP, the stronger is the quantization of the sequence and the lower is the resulting video quality. As the value of the QP parameter any integer between 0 and 51 can be selected. The QP can either be a constant integer or - using rate control - automatically dynamically adjusted to match a previously selected bitrate.\\
For the evaluation, the 'Foreman' (CIF, 30 Hz), 'Crew' (4CIF, 60 Hz) and 'Pedestrian Area' (720p, 25 Hz) sequence are each encoded with a single layer and constant QPs of 0, 10, 20, 30, 40 and 50. These sequences are used as they provide a wide range of different motion and spatial details. All other encoder settings are left at standard values. So, for each sequence six videos are made and evaluated in the test.

\subsubsection{Optimal quantization parameter test}By using the results of the quantization parameter test and the filesizes of the encoded sequences, a QP range in which the optimal ratio of filesize and visual quality is achieved is pinpointed. To get exact results, each of the evaluated sequences ('Foreman', 'Crew', 'Pedestrian Area') is additionally encoded with 9 different QP settings ranging from 31 to 39 in single steps. These impaired sequences are then assessed.\\
With the resulting quality scores and related filesizes for each QP setting, the exact location of the optimal quantization parameter is calculated.

\subsubsection{CGS / MGS test}In the coarse grain scalability (CGS) / medium grain scalability (MGS) test, the impact of MGS on the video quality is assessed in comparison with CGS coding. To do so, the three sequences already used previously in the quantization parameter test ('Foreman', 'Crew', 'Pedestrian Area') are encoded with two layers (Base layer (BL) and enhancement layer (EL)). In CGS mode, only these two layers - using SNR scalability - could be extracted, while the sequence encoded with MGS additionally offered 4$\times$4 MGS vectors to dynamically adjust to changing bandwidth needs. Except for the two layers, the standard encoding settings are employed.\\
During the test, three different bitrates are compared. For each bitrate setting, a video stream is extracted out of the SVC file. The three sequences of each test video are then evaluated by the test subjects to determine if there is an impact of MGS on perceived quality in this setting and how big it is. The encoding bitrates for each layer (base layer bitrate (BLB), enhancement layer bitrate (ELB) and extraction bitrates (EBs)) are:

\begin{table}[H]
\begin{center}
\begin{tabular}{| l | c |}\hline
 &\textbf{'Foreman'}\\\hline
\textbf{BLB}&$262$\\\hline
\textbf{ELB (CGS)}&$1077$\\\hline
\textbf{ELB (MGS)}&$1289$\\\hline
\textbf{EBs}&$500, 1000, 1500$\\\hline
\end{tabular}
\caption{Encoding and extraction bitrates used in the CGS / MGS test.}
\label{tab:BITRATES_CGS_MGS_CIF}
\end{center}
\end{table}

\begin{table}[H]
\begin{center}
\begin{tabular}{| l | c |}\hline
 &\textbf{'Crew'}\\\hline
\textbf{BLB}&$2409$\\\hline
\textbf{ELB (CGS)}&$10707$\\\hline
\textbf{ELB (MGS)}&$10101$\\\hline
\textbf{EBs}&$3000, 7000, 11000$\\\hline
\end{tabular}
\caption{Encoding and extraction bitrates used in the CGS / MGS test.}
\label{tab:BITRATES_CGS_MGS_4CIF}
\end{center}
\end{table}

\begin{table}[H]
\begin{center}
\begin{tabular}{| l | c |}\hline
 &\textbf{'Pedestrian Area'}\\\hline
\textbf{BLB}&$1353$\\\hline
\textbf{ELB (CGS)}&$4719$\\\hline
\textbf{ELB (MGS)}&$5644$\\\hline
\textbf{EBs}&$3000, 4500, 6000$\\\hline
\end{tabular}
\caption{Encoding and extraction bitrates used in the CGS / MGS test.}
\label{tab:BITRATES_CGS_MGS_HD}
\end{center}
\end{table}

As MGS encoding introduces an additional overhead to the SVC stream due to the availability of multiple MGS vectors the ELB bitrates of CGS and MGS sequences differ.

\subsubsection{Best extraction path test}As the different video streams embedded in a SVC bitstream are arranged in a spatio-temporal cube, the best extraction path test is conducted to determine which of the video streams is perceived as the optimal one for a given bitrate in terms of visual quality.\\
To achieve this, the unimpaired original 4CIF sequences are encoded in three spatial (QCIF, CIF, 4CIF) and four temporal (7.5 Hz, 15 Hz, 30 Hz, 60 Hz) resolutions each. The QP of each layer is adjusted to match the target filesize of 1000 KB. The resulting 12 impaired sequences are compared in the evaluation.\\
The outcome of the best extraction path test shows which of the three kinds of impairments (spatial, temporal or SNR) has the biggest impact on perceived quality and, as a result, if there is an extraction path which is can generally be recommended or if the results are highly dependent on the content of the encoded sequence.

\subsubsection{Packet loss test:}The SVC CODEC's scalable features are most advantageous in streaming environments, especially the Internet, where the available bandwidth of each client differs significantly. Besides the bandwidth, the response time is also an important aspect of the connection. To provide low delays, multimedia servers nearly exclusively rely on connections over RTP, which is based on UDP \cite{RTP1} \cite{RTP2}. While providing small delays and timestamps (among other features), this protocol has the severe disadvantage that no error correction is supported. The result is that transmissions over error-prone channels manifest in visual impairments of the streamed video file.\\
To test the behavior of the SVC CODEC in the case of errors, an error recovery test was conducted using the 'Foreman' sequence. The file was then encoded using:

\begin{itemize}
\item{The standard encoding settings, containing the following bitstreams after encoding:
\begin{itemize}
\item{Layer-ID 0: 352$\times$288, 7.5 Hz, 180.9 kbps}
\item{Layer-ID 1: 352$\times$288, 15 Hz, 216.7 kbps}
\item{Layer-ID 2: 352$\times$288, 30 Hz, 257.7 kbps}
\end{itemize}}
\item{Two layers with spatial scalability (QCIF \& CIF resolution), containing the following bitstreams after encoding:
\begin{itemize}
\item{Layer-ID 0: 176$\times$144, 7.5 Hz, 60.2 kbps}
\item{Layer-ID 1: 176$\times$144, 15 Hz, 77.2 kbps}
\item{Layer-ID 2: 176$\times$144, 30 Hz, 94.5 kbps}
\item{Layer-ID 3: 352$\times$288, 7.5 Hz, 240.8 kbps}
\item{Layer-ID 4: 352$\times$288, 15 Hz, 294.2 kbps}
\item{Layer-ID 5: 352$\times$288, 30 Hz, 353.0 kbps}
\end{itemize}}
\item{Two layers with SNR scalability (QP 36 \& QP 26), containing the following bitstreams after encoding:
\begin{itemize}
\item{Layer-ID 0: 352$\times$288, 7.5 Hz, 104.1 kbps}
\item{Layer-ID 1: 352$\times$288, 15 Hz, 136.8 kbps}
\item{Layer-ID 2: 352$\times$288, 30 Hz, 175.4 kbps}
\item{Layer-ID 3: 352$\times$288, 7.5 Hz, 696.4 kbps}
\item{Layer-ID 4: 352$\times$288, 15 Hz, 845.8 kbps}
\item{Layer-ID 5: 352$\times$288, 30 Hz, 1010.1 kbps}
\end{itemize}}
\end{itemize}

As temporal scalability is already present in the SVC file encoded with standard settings, this feature was not evaluated separately.\\
During the evaluation, packet loss was simulated using the packet loss simulation tool (PacketLossSimulatorStatic.exe), which is included in the current SVC build. Further details concerning this tool are provided in \cite{softman}. The tested sequences were exposed to four levels of error: 3\%, 5\%, 10\% and 20\%. The impact of the errors on the video quality and the resulting impairments were then evaluated for each single bitstream.

\subsection{Comparison of MPEG-4 SVC to MPEG-4 AVC/ASP}

\subsubsection{Quality comparison test}During the quality comparison test, nine test sequences are encoded with the three evaluated CODECs Xvid, x264 and SVC. The CIF sequences are encoded with 200 kbps, the 4CIF and HD sequences with 1000 kbps.\\
In the subjective assessment, the experts are then asked to evaluate the sequences: In each test, the subject is first shown the uncompressed reference sequence. After that, the three impaired versions of the same sequence compressed with the three evaluated CODECs are compared to the original.\\
During the objective evaluation, the three impaired sequences encoded with the tested CODECs of each sequence are compared to each other.\\
The results of this test show which of the three CODECs produces the best quality in mean and if or how great the bitrate and resolution impact the quality of each CODEC.

\subsubsection{Encoding speed test}In the encoding speed test, the time of each CODEC to encode a given sequence is measured. For this evaluation the standard encoder settings are employed. For the encoding process, three sequences ('Foreman', 'Crew' and 'Pedestrian Area') with different resolutions and a duration of 10 seconds each are used.\\
The sequences are looped 3 times before the encoding process with Xvid or x264 to reduce measuring accuracies. This is necessary as the encoding times with these CODECs are very short for the non-looped sequences. SVC encoding in contrast is unproblematic in this respect due to its lower encoding speed. Additionally to the testing of all CODECs using their standard settings, the speed of the x264 CODEC is also evaluated when the parameter 'Threads' is reduced to '1', to investigated the impact multithreading has on its encoding speed.\\

\section{Results}
\label{sec:results}

\subsection{MPEG-4 SVC stand-alone test}
First, the results from different tests regarding the SVC options are compared. It has to be mentioned that some tests could only be performed using objective metrics as the differences in quality are too small to be evaluated subjectively.

\subsubsection{Encoding parameter test}

\paragraph{Motion estimation}
As the following paragraphs show, motion estimation has only little impact on visual quality or bitrate. However, these options have a high influence on encoding speed.\\
First, the results of the comparison of different block matching metrics are presented. As the results in table \ref{tab:motion_estimation} show, the effect of different block matching metrics on Y-PSNR are rather small but can have a major impact on encoding time.

\begin{table}[H]
\scriptsize{
\centering
\begin{tabular}{|l|l|r|r|}
\hline
\textbf{FullPel} & \textbf{SubPel} & \textbf{$\Delta$Enc. time} & \textbf{$\Delta$Y-PSNR}\\
\hline
SAD & SAD & $\pm 0{.}0000\%$ & $\pm 0{.}0000\%$\\
SAD & SSE & $+4{.}4902\%$ & $-0{.}2384\%$\\
SAD & HADAMARD & $+8{.}0449\%$ & $+0{.}2517\%$\\
SSE & SAD & $+4{.}3966\%$ & $-0{.}0086\%$\\
SSE & SSE & $+9{.}6352\%$ & $-0{.}2442\%$\\
SSE & HADAMARD & $+9{.}8223\%$ & $+0{.}2282\%$\\
HADAMARD & SAD & $+143{.}5921\%$ & $+0{.}0487\%$\\
HADAMARD & SSE & $+153{.}0402\%$ & $-0{.}2012\%$\\
HADAMARD & HADAMARD & $+146{.}9598\%$ & $+0{.}2785\%$\\
SAD-YUV & SAD & $+33{.}9570\%$ & $+0{.}0036\%$\\
SAD-YUV & SSE & $+36{.}9504\%$ & $-0{.}2328\%$\\
SAD-YUV & HADAMARD & $+29{.}5603\%$ & $+0{.}2563\%$\\
\hline
\end{tabular}
\caption{Impact of different block matching metrics.}
\label{tab:motion_estimation}
}
\end{table}

Another interesting fact is that Y-PSNR is mostly independent from the FullPel block matching metric. Concerning the SubPel metric, 'HADAMARD' always reaches the highest Y-PSNR values, while 'SSE' reaches the lowest, regardless of the used FullPel metric. On the other hand, the required processing time depends primarily on the chosen FullPel metric, 'SAD' is the best choice here. 'HADAMARD' is a poor choice for FullPel as it leads to a significant increase in encoding time, however Y-PSNR does not profit much from it. Concluding, a combination of 'SAD' as FullPel and 'HADAMARD' as SubPel metric can be recommended.\\
The next aspect analyzed is the effect of different block matching algorithms. Among the results, the most striking point is that block search has little impact on visual quality but leads to a very large increase in encoding time, as table \ref{tab:me_algo} shows.

\begin{table}[H]
\centering
\begin{tabular}{|l|r|r|r|}
\hline
\textbf{Algorithm} & \textbf{$\Delta$Enc. time} & \textbf{$\Delta$Filesize} & \textbf{$\Delta$Y-PSNR}\\
\hline
Block search & $\pm 0{.}0000\%$ & $\pm 0{.}0000\%$ & $\pm 0{.}0000\%$\\
Fast search & $-96{.}0624\%$ & $+0{.}8825\%$ & $+0{.}0066\%$\\
\hline
\end{tabular}
\caption{Impact of different block matching algorithms.}
\label{tab:me_algo}
\end{table}

Even when normalizing the bitrates, the advantage of block search in Y-PSNR values increases only very slightly. but the gain is still very small compared to the increase in encoding time of $+2439{.}65\%$. In common scenarios, 'Fast search' is the much more feasible choice among the two algorithms.\\
An interesting fact is that the processing time needed for block search is independent from the motion present in the source video, this is different for the fast search algorithm. The exact values are given in table \ref{tab:fast_dep_tab}.

\begin{table}[H]
\centering
\begin{tabular}{|l|c|}
\hline
\textbf{Sequence} & \textbf{Encoding time per frame [s]}\\
\hline
'Bus' & $0{.}9333$\\
'Football' & $1{.}1846$\\
'Foreman' & $0{.}7767$\\
\hline
\end{tabular}
\caption{Dependence of fast search on source videos.}
\label{tab:fast_dep_tab}
\end{table}

This is due to the early break criteria present in most fast search algorithms. As soon as the chosen block matching metric value falls under a certain threshold for the considered block candidate any further evaluation of candidates for this block is omitted. This explains why fast search takes most time for high motion ('Football' sequence) and least for low motion content ('Foreman' sequence).

\subsubsection{Quantization parameter test}

\begin{figure}[H]
\centering
\includegraphics[width=0.45\textwidth]{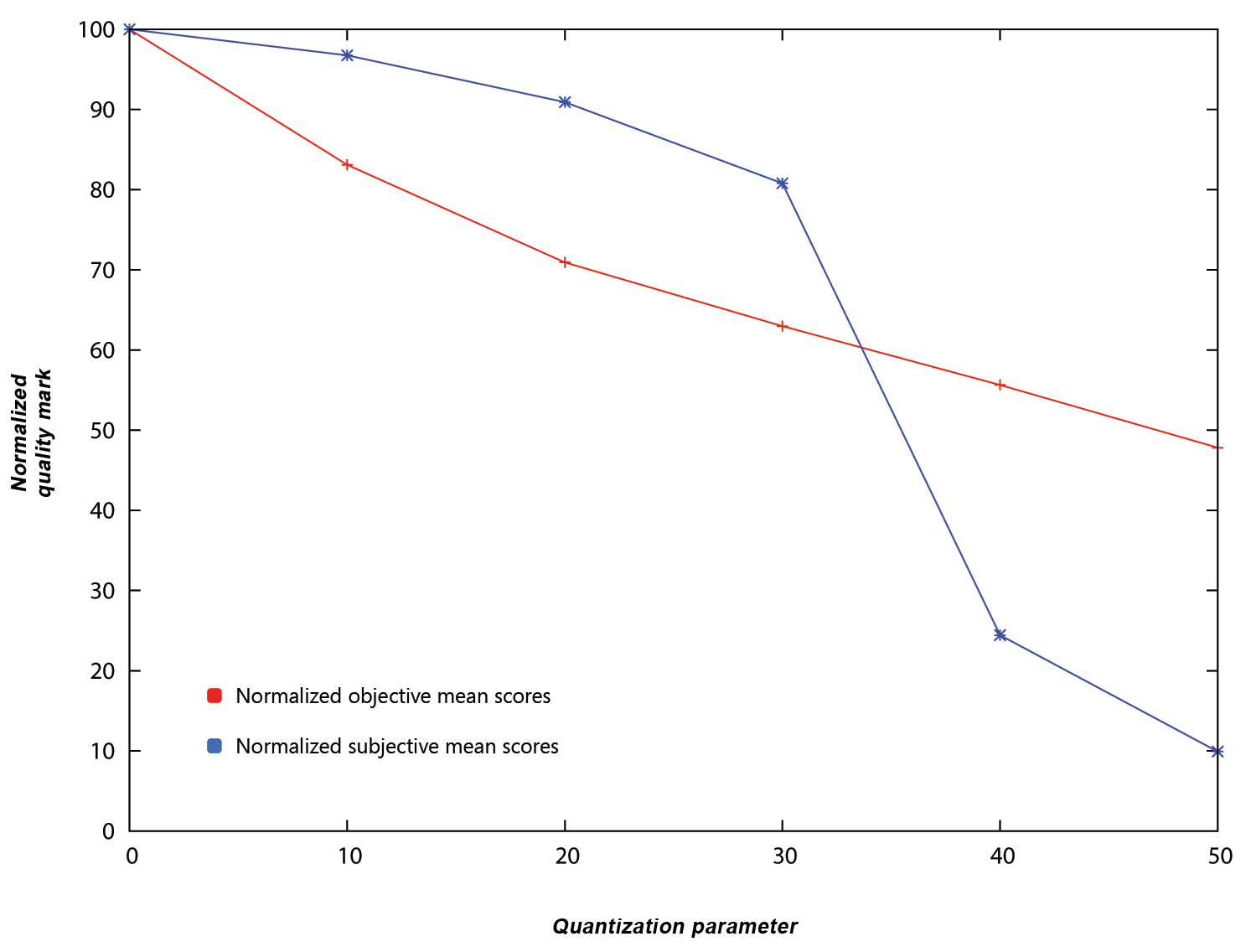}
\caption{Normalized average marks of the objective and subjective quality in the quantization parameter test.}
\label{fig:QP_TEST}
\end{figure}

Figure \ref{fig:QP_TEST} shows a comparison of objective and subjective quality scores obtained in the quantization parameter test. As easily visible, both scores differ significantly: While the objective score degrades almost linearly with the rising QP value, the subjective score shows very little quality impairment up to a QP value of 30, but then quickly falls to a relative score of about 25\% at QP 40.

This test shows that there is a significant gap between PSNR and subjectively perceived quality. Apparently a certain amount of loss in high frequency information does not impair perceived quality much, but of course this loss is already picked up by the PSNR calculation. The intersection of both graphs is at about QP 33, both scores reach about 61\% relative quality there.

\subsubsection{Optimal quantization parameter test}

\begin{figure}[H]
\centering
\includegraphics[width=0.45\textwidth]{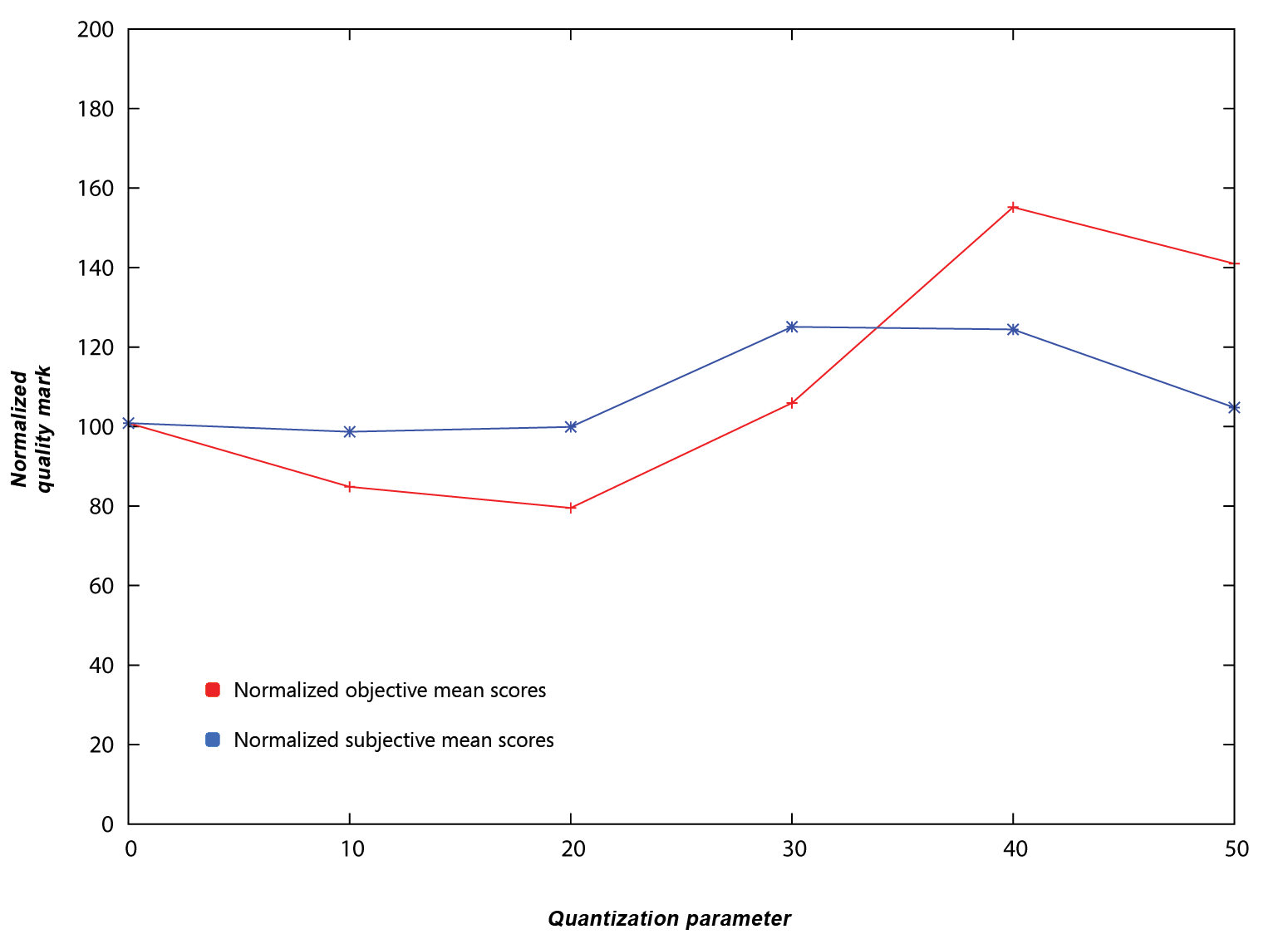}
\caption{Approximation of the optimal objective and subjective quantization parameter.}
\label{fig:SYN_OPTIQP_ROUGH}
\end{figure}

In both objective and subjective testing, an optimal choice for the quantization parameter is derived. In this evaluation, the optimal QP setting is located where the best relation of filesize and visual quality is present: $MAX(normalize(\frac{1}{Bitrate}) + normalize(Visual~Quality))$.\\
First, the whole range of the possible quantization parameter settings is tested in steps of $10$ points. The results are shown in figure \ref{fig:SYN_OPTIQP_ROUGH}.

\begin{figure}[H]
\centering
\includegraphics[width=0.45\textwidth]{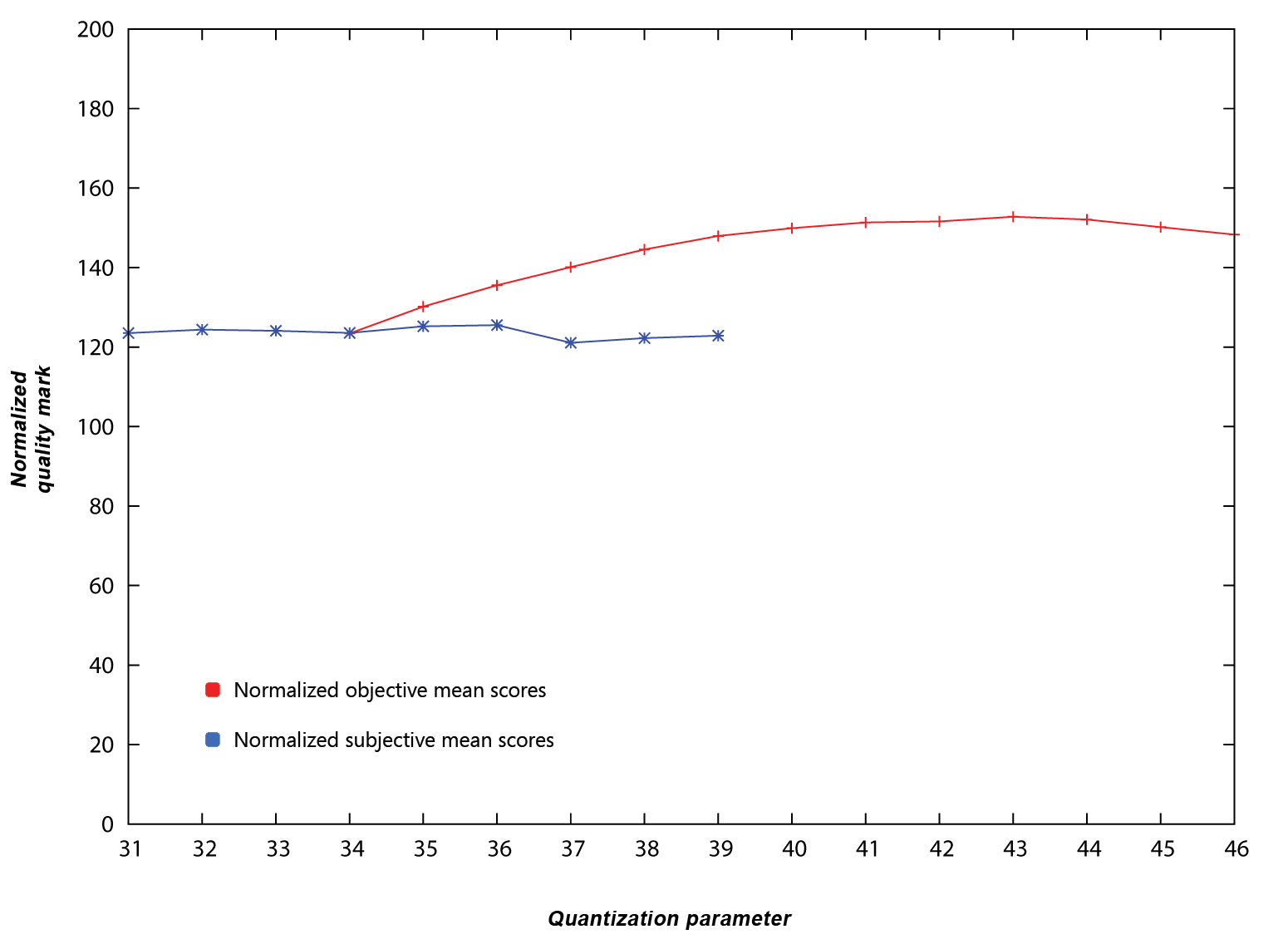}
\caption{Optimal objective and subjective quantization parameter.}
\label{fig:SYN_OPTIQP}
\end{figure}

Using this result, a fine granular search for the optimum quantization parameter value is conducted. Figure \ref{fig:SYN_OPTIQP} shows the significant differences between objective and subjective results.

While objective score reaches its maximum at QP $43$, the subjective maximum is at QP $36$. The reason for this discrepancy is the high subjective quality loss in the region between QP $30$ and $40$, while Y-PSNR values show only moderate decreases in this region.

\subsubsection{CGS / MGS test}

\begin{figure}[H]
\centering
\includegraphics[width=0.43\textwidth]{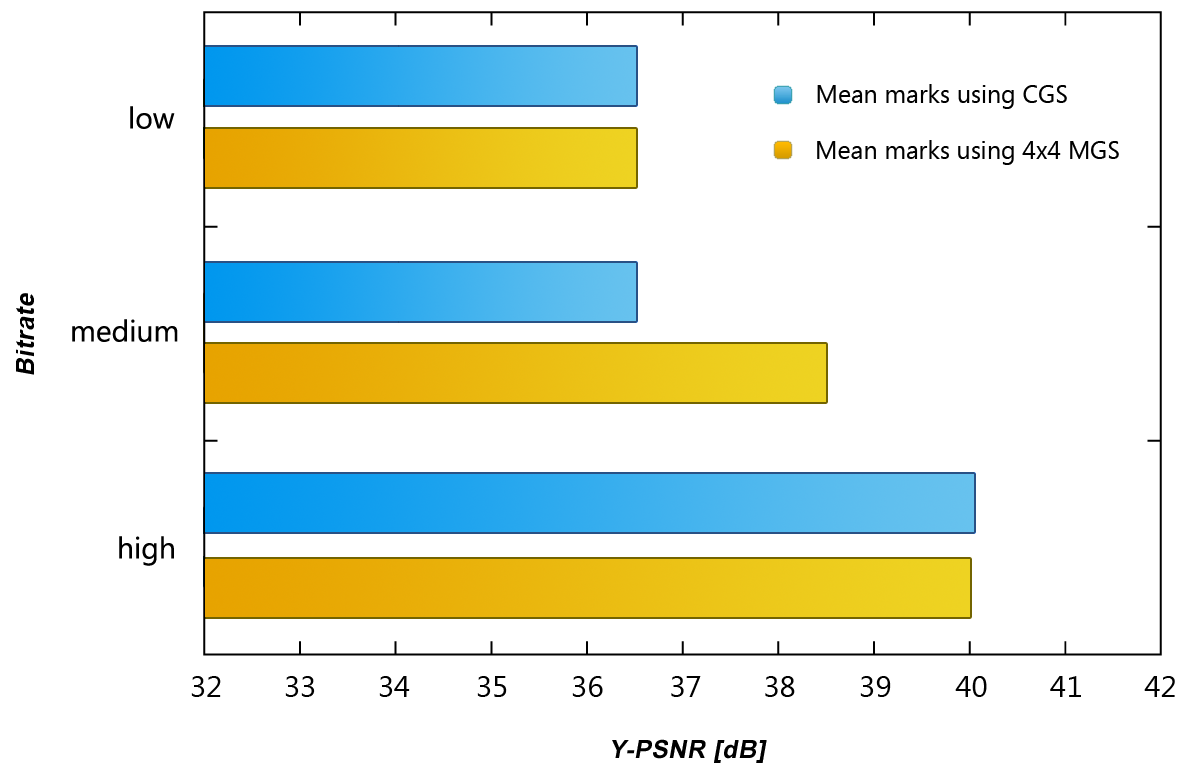}
\caption{Objective mean scores for CGS and MGS coded sequences.}
\label{fig:cgs_mgs_obj}
\end{figure}

\begin{figure}[H]
\centering
\includegraphics[width=0.43\textwidth]{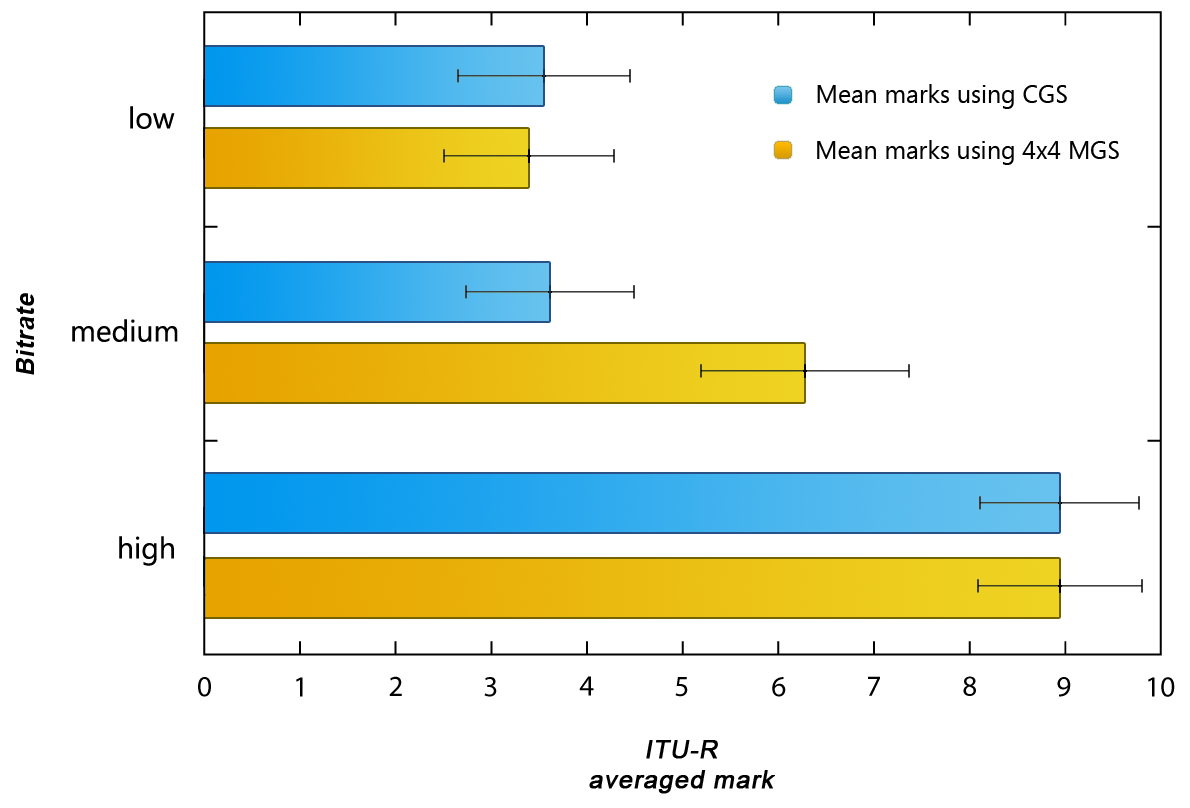}
\caption{Subjective mean scores for CGS and MGS coded sequences.}
\label{fig:cgs_mgs_sub}
\end{figure}

The CGS / MGS test show similar results in both objective and subjective evaluation. At bitrates between the two SNR layers, MGS encoding can lead to a significant increase in quality. Figure \ref{fig:cgs_mgs_obj} and \ref{fig:cgs_mgs_sub} show the relative gain of objective and subjective values using $4$ MGS vectors.

\begin{figure}[H]
\centering
\includegraphics[width=0.45\textwidth]{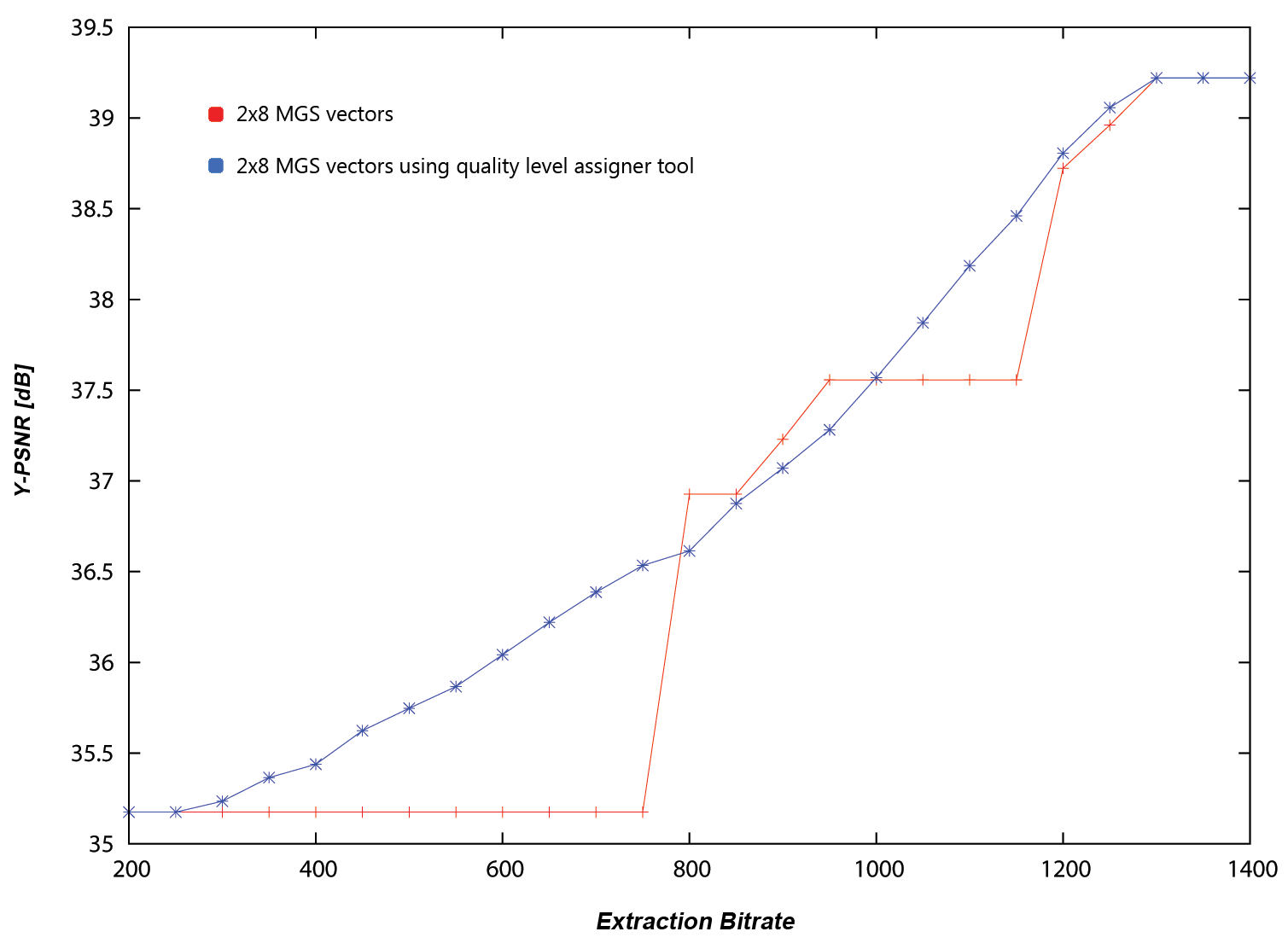}
\caption{Comparison of the 'Foreman' sequence using 2$\times$8 MGS
vectors with and without the quality level assigner tool.}
\label{fig:qla}
\end{figure}

As the objective tests show, the quality level assigner tool can be
used to achieve an almost linear PSNR increase with a low number of
MGS vectors, which can be seen in figure \ref{fig:qla}.

\subsubsection{Best extraction path test}

While the results of the objective best extraction path assessment showed the best PSNR values for sequences encoded in 4CIF resolution and 30 / 60 Hz, in subjective testing, in contrast, especially the bitstream using the highest possible spatial and temporal level is rated very poor. This finding matches with the ones previously mentioned in the quantization parameter test, where the subjective quality ratings suddenly drops between QP 30 and 40, whereas the objective scores scaled almost linearly throughout the whole QP range. Because the QP had to be adjusted higher than 40 when 4CIF resolution and a framerate of 60 Hz is used in the 'Harbour' and 'Crew' sequence, the corresponding quality scores are much lower in the subjective test than in the objective one.
In the following figures, the numbers from 1 to 12 indicate the visual quality of each selectable bitstream, where 1 is the best and 12 the worst rating.

\begin{figure}[H]
\centering
\includegraphics[width=0.45\textwidth]{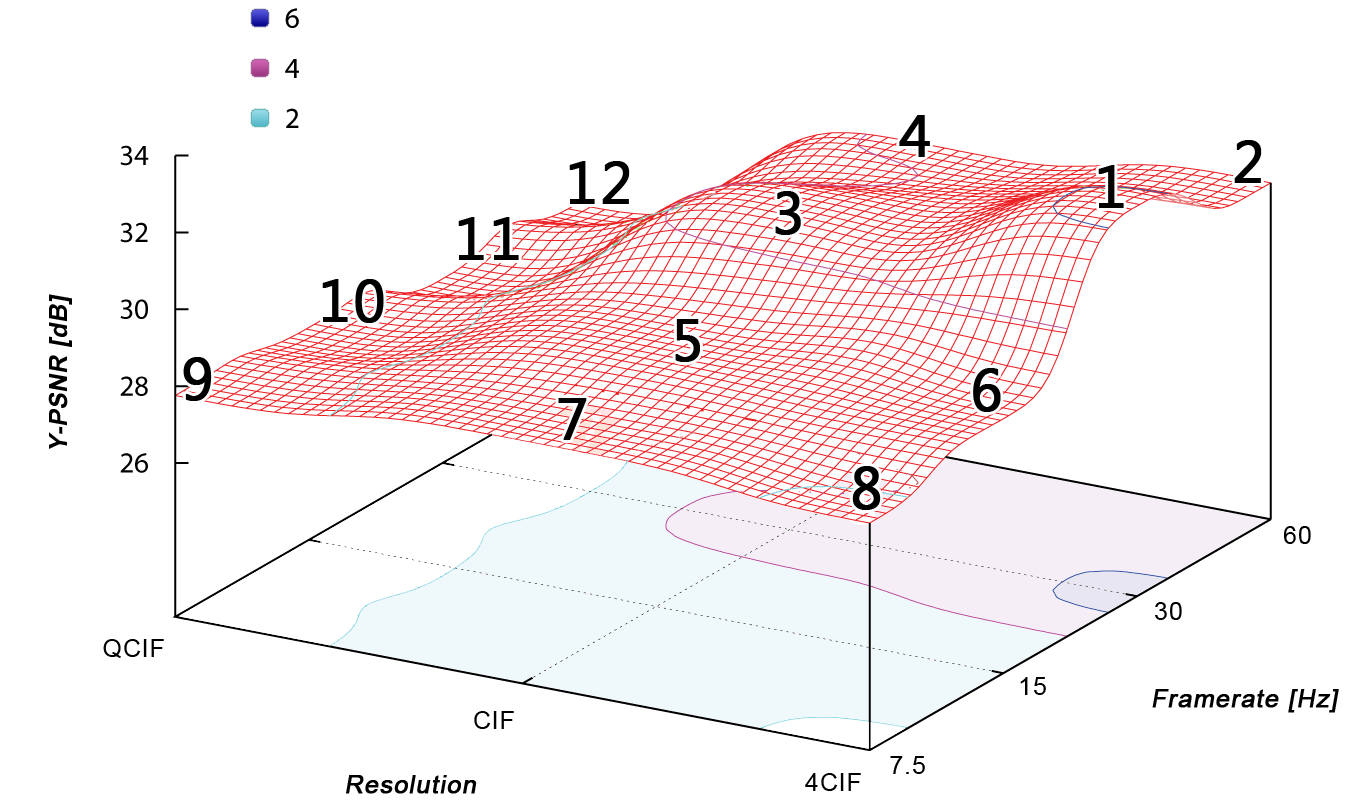}
\caption{Objective and subjective quality marks for different framerates and resolutions.}
\label{fig:EXPATH_OBJ}
\end{figure}

\begin{figure}[H]
\centering
\includegraphics[width=0.45\textwidth]{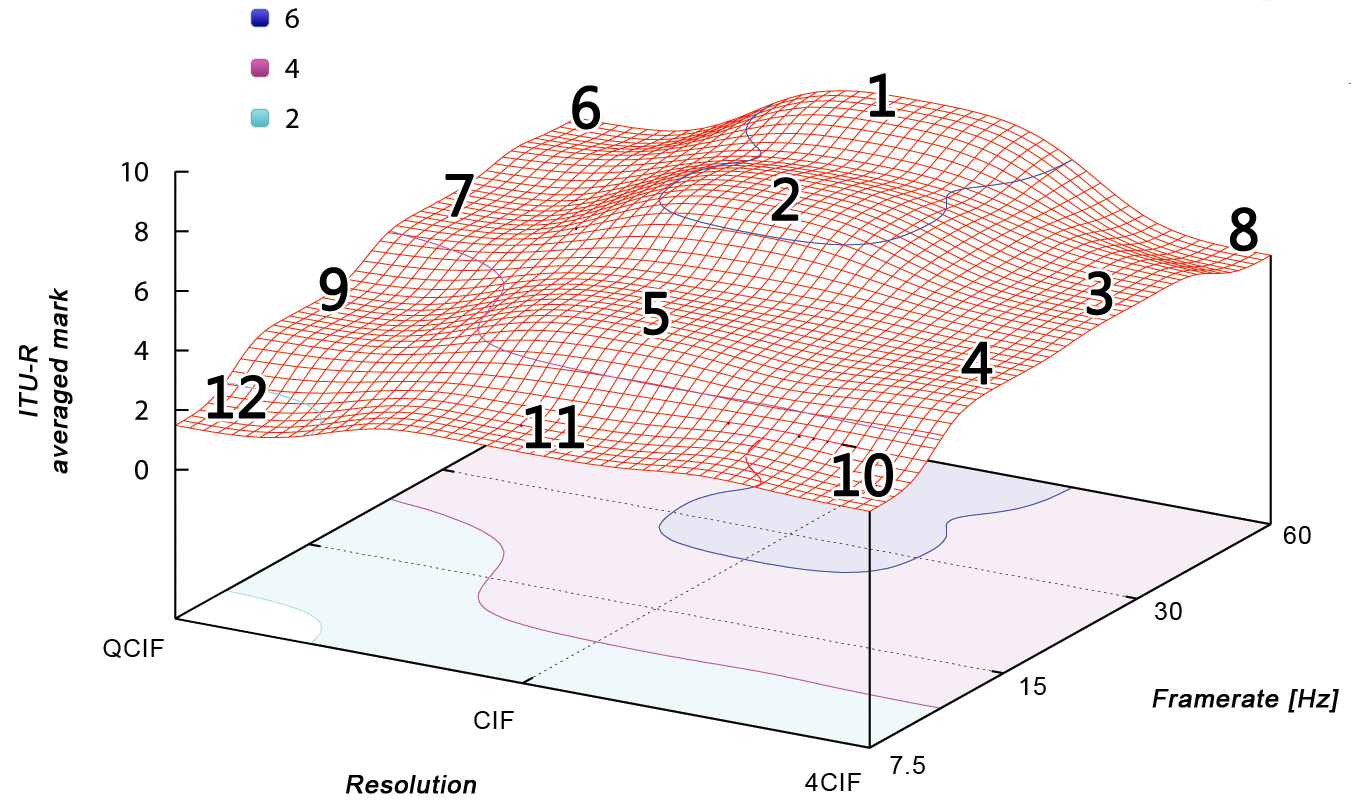}
\caption{Subjective quality marks for different framerates and resolutions.}
\label{fig:EXPATH_SUB}
\end{figure}

Apart from that, it is additionally visible that QCIF resolution, as well as all streams encoded with 7.5 Hz framerate received very low scores in both test runs. As a result, the selection of the lowest spatial and/or temporal resolution should be avoided as far as possible.\\
The differences of the objective and subjective testings are visualized in the following figure, where the difference of the results of both assessments is calculated. Scores higher than 0 show that a video stream received a higher rating in the objective evaluation, while negative marks indicate that subjective rating is higher than the objective.

\begin{figure}[H]
\centering
\includegraphics[width=0.45\textwidth]{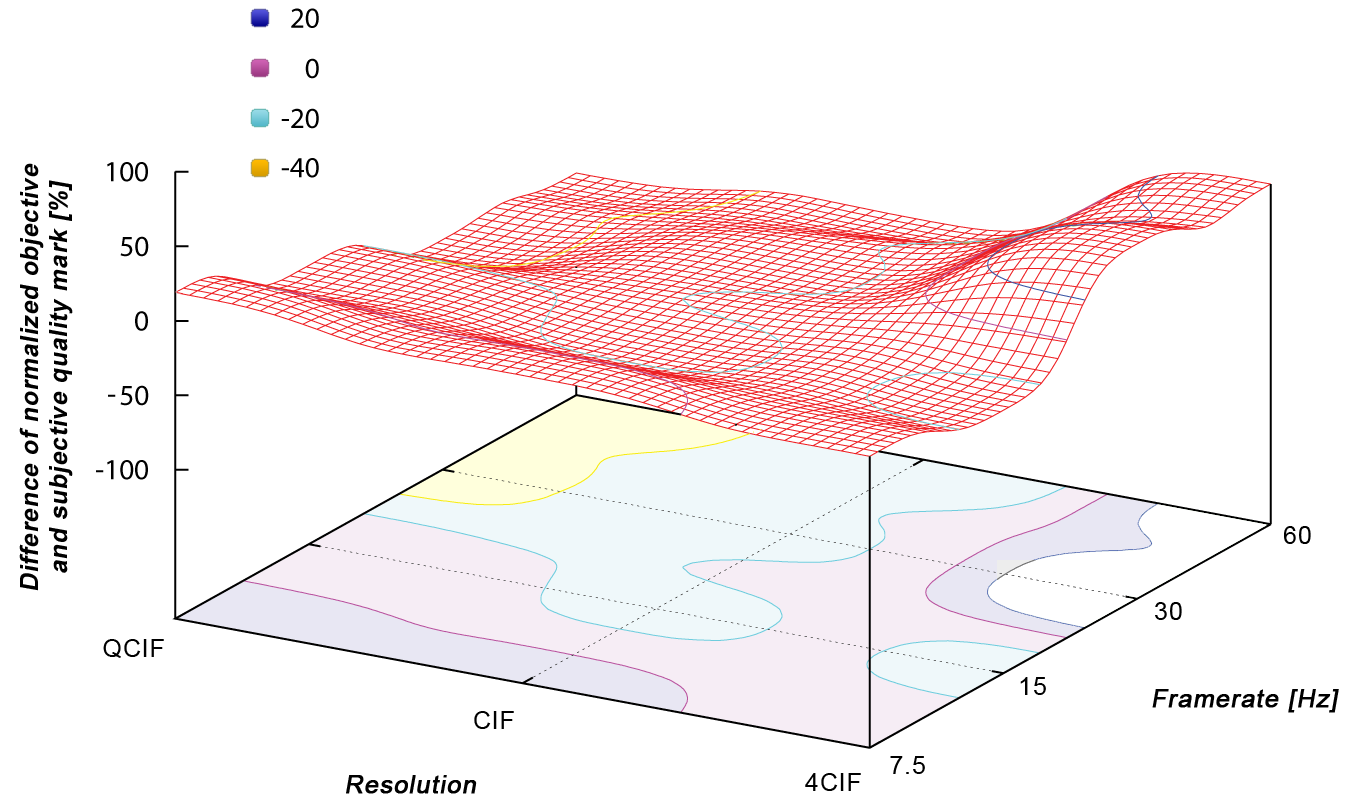}
\caption{Comparison of mean objective and subjective quality marks for different framerates and resolutions.}
\label{fig:EXPATH_MEAN}
\end{figure}


\subsubsection{Packet loss test:} The results of the error recovery test were different when comparing a single layer encode to one using multiple layers. The detailed results are listed below for each encoding setting.\\

\textbf{Standard encoding parameters:} When the standard encoding parameters were used, the decodeable video duration was shortened linearly with the amount of packet errors. That applied to both bitstreams with high as well as low layer-IDs in this case. The image artifacts manifested in an increasing amount of blocking and color distortions the higher the packet error rate was adjusted.

\begin{table}[H]
\begin{center}
\begin{tabular}{| c | c | c | c |}\hline
\textbf{Error rate}&\textbf{Layer}&\textbf{\# decodable}&\textbf{ITU-R mark}\\\hline
&0&71&2.42\\
3\%&1&143&3.79\\
&2&289&5.96\\\hline

&0&70&0.75\\
5\%&1&142&1.53\\
&2&280&5.79\\\hline

&0&64&0.80\\
10\%&1&130&2.56\\
&2&265&5.42\\\hline

&0&57&0.78\\
20\%&1&116&1.14\\
&2&235&2.34\\\hline
\end{tabular}
\caption{Impact of packet errors on the decodeable video duration and subjective quality.}
\label{tab:DURATION_STANDARD_ERROR}
\end{center}
\end{table}

The duration of the unimpaired sequences depends on the used framerate, therefore the full number of frames are: 300 for 30 Hz (layer-ID 2), 150 for 15 Hz (layer-ID 1) and 75 for 7.5 Hz (layer-ID 0).\\
This applies to the following test scenarios too, only the number of layer-IDs is doubled because of the additional enhancement layer. The full duration is therefore: 300 for layer-ID 2 \& 5, 150 for layer-ID 1 \& 4 and 75 for layer-ID 0 \& 3.\\

\textbf{Spatial scalability:} The video file with spatial scalable layers showed a different behavior during the assessment. Even low error rates had a serious impact on the bitstream with layer-IDs below 3. When the error rate reached 5\%, layer-ID 0 was not decodeable, by 10\%, all three smallest bitstreams were not viewable anymore. The layer-IDs bigger than 3 were still decently viewable up to a error rate of 5\%, then the decodeable duration of the sequence was shortened to 55.83\% in mean. Additionally, severe color distortions were already present in files that were processed with 5\% error rate.

\begin{table}[H]
\begin{center}
\begin{tabular}{| c | c | c | c |}\hline
\textbf{Error rate}&\textbf{Layer}&\textbf{\# decodable}&\textbf{ITU-R mark}\\\hline
&0&23&0.91\\
&1&58&2.58\\
&2&22&1.82\\
3\%&3&64&1.92\\
&4&135&5.37\\
&5&280&6.83\\\hline

&0&---&---\\
&1&9&0.36\\
&2&31&0.92\\
5\%&3&61&0.99\\
&4&130&1.21\\
&5&236&1.32\\\hline

&0&---&---\\
&1&---&---\\
&2&---&---\\
10\%&3&67&2.16\\
&4&86&3.48\\
&5&163&3.85\\\hline

&0&---&---\\
&1&---&---\\
&2&---&---\\
20\%&3&50&0.83\\
&4&107&0.49\\
&5&110&0.59\\\hline
\end{tabular}
\caption{Impact of packet errors on the decodeable video duration and subjective quality.}
\label{tab:DURATION_SPATIAL_ERROR}
\end{center}
\end{table}

\textbf{SNR scalability:} The SNR-scalability test sequences suffered similar duration shortenings as the spatial scalable ones. Also the problem that the three lower bitstreams are not decodeable when the error rate reaches 10\% reappears during the SNR-scalability test. The visual impairments however manifested in a mixture of blocking artifacts, reference frame errors and complete picture losses lasting for one frame.

\begin{table}[H]
\begin{center}
\begin{tabular}{| c | c | c | c |}\hline
\textbf{Error rate}&\textbf{Layer}&\textbf{\# decodable}&\textbf{ITU-R mark}\\\hline
&0&27&1.31\\
&1&62&6.25\\
&2&26&1.89\\
3\%&3&64&2.93\\
&4&135&6.81\\
&5&293&6.25\\\hline

&0&2&0.22\\
&1&13&0.40\\
&2&35&0.48\\
5\%&3&61&0.55\\
&4&114&0.76\\
&5&152&0.76\\\hline

&0&---&---\\
&1&---&---\\
&2&---&---\\
10\%&3&67&1.88\\
&4&75&2.85\\
&5&117&2.87\\\hline

&0&---&---\\
&1&---&---\\
&2&---&---\\
20\%&3&49&0.55\\
&4&123&0.85\\
&5&48&0.61\\\hline
\end{tabular}
\caption{Impact of packet errors on the decodeable video duration and subjective quality.}
\label{tab:DURATION_SNR_ERROR}
\end{center}
\end{table}

The tables~\ref{tab:DURATION_STANDARD_ERROR},~\ref{tab:DURATION_SPATIAL_ERROR} and~\ref{tab:DURATION_SNR_ERROR} show the encoding settings, error rates, layer-IDs, number of decodeable video frames and the corresponding ITU averaged mark.\\

\begin{figure}[H]
\centering
\includegraphics[width=0.46\textwidth]{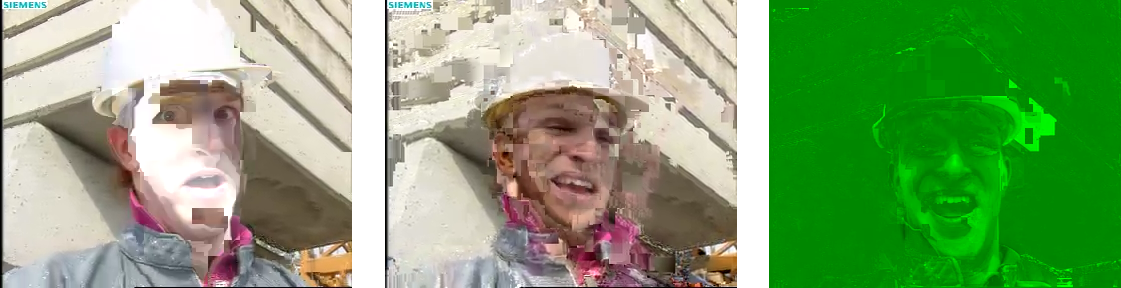}
\caption{Different artifacts due to packet loss.}
\label{fig:LOSS-ERROR}
\end{figure}

In conclusion, when comparing the subjective quality marks, it is visible that the encode using standard settings in mean suffers less of the packet loss than any bitstream with multiple scalable layers. The quality developed as expected in this test case: With increasing error rate, the visual quality decreased steadily.\\
If scalable layers were present, especially the quality of the bitstreams with small layer-IDs (0 - 2) suffered severely under the packet loss, making them completely undecodeable if 10\% or more error rate was selected.\\
An interesting development could be observed in the highest layer-ID (5): While the visual quality was nearly on par in all encoding scenarios when an error rate of 3\% was used, the scalable encoded sequences did - in contrast to the standard settings encode - not show a clear negative trend, which would be expected when applying increasing error rates. Instead, the perceived quality of 3\% and 10\% error rate were much higher than the one of 5\% and 20\% error rate.\\
This is surprising, as the decodeable duration of the sequences with 10\% error rate is lower than that of the ones with 5\%. The main reason for the low scores is most likely the overall loss of color information in both scalable encodes with 5\% error rate, which is not present in the encode with 10\% errors.
An impression of some typical error patterns in the sequences is given in figure~\ref{fig:LOSS-ERROR}.

\subsection{Comparison of MPEG-4 SVC to MPEG-4 AVC/ASP}

\subsubsection{Quality comparison test}

When looking at the quality comparison test, basically similar results could be observed in both subjective and objective testing. The overall visual quality of the three tested CODECs in the evaluated scenarios leads to the following ranking:

\begin{figure}[H]
\centering
\includegraphics[width=0.45\textwidth]{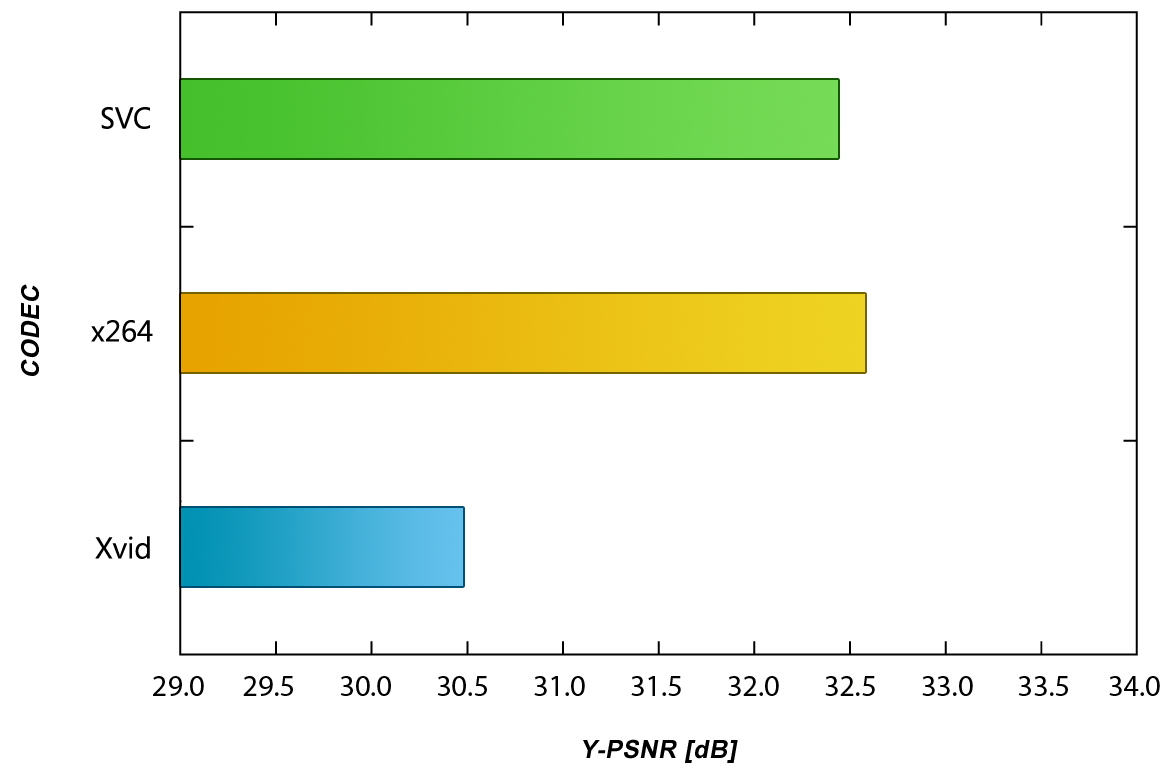}
\caption{Objective quality results of the quality comparison test.}
\label{fig:codec_comp_obj}
\end{figure}

\begin{figure}[H]
\centering
\includegraphics[width=0.45\textwidth]{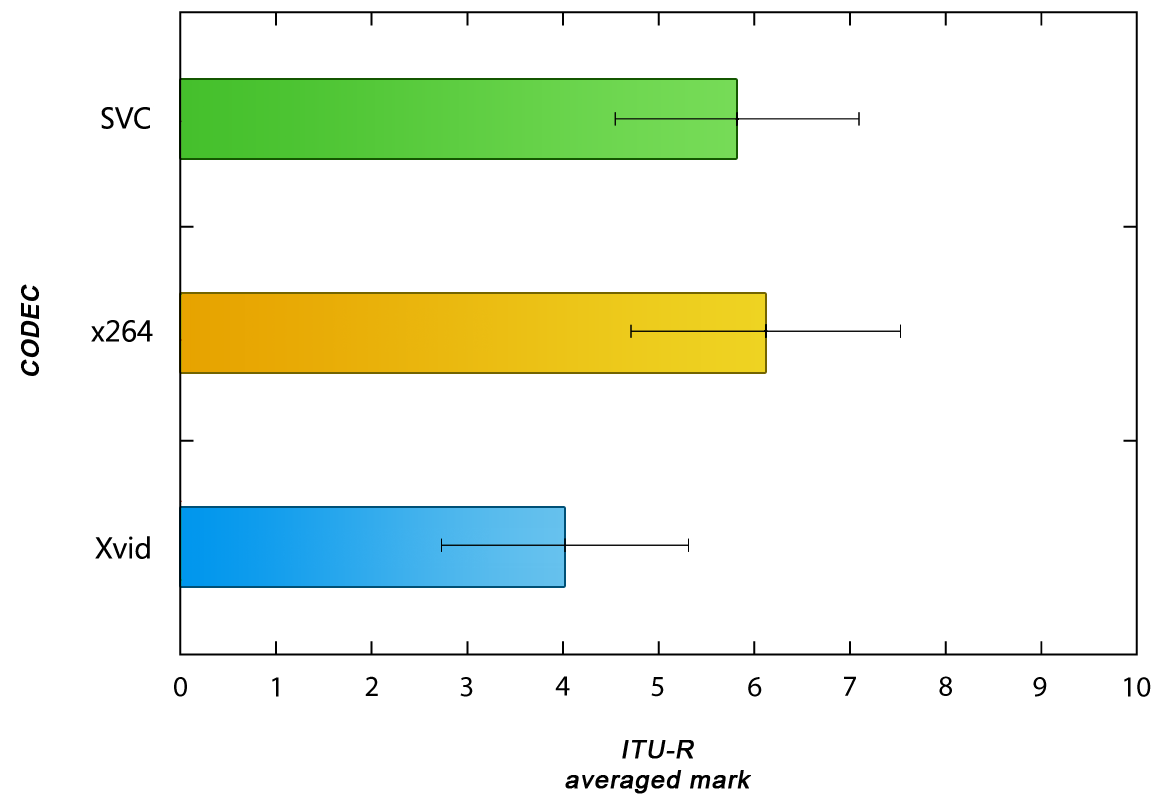}
\caption{Subjective quality results of the quality comparison test.}
\label{fig:codec_comp_sub}
\end{figure}

These results are not surprising,   considering Xvid is by far the oldest CODEC in the comparison, while the performance of MPEG-4 SVC - if the standard settings are employed - is expected to be similar to that of x264, because SVC is directly based on the MPEG-4 AVC standard. Small differences in quality are attributable to the higher technical maturity and optimizations of the x264 CODEC in this case.\\
The only significant difference is that quality variations show a higher amplitude in the subjective evaluation than in the objective one. This becomes especially visible when looking at the results of the Xvid CODEC, where the objective still scored 93.5\%, while the subjective mark is only 65.2\%. The same phenomenon is already observeable in the CGS / MGS assessment.\\

\begin{figure}[H]
\centering
\includegraphics[width=0.45\textwidth]{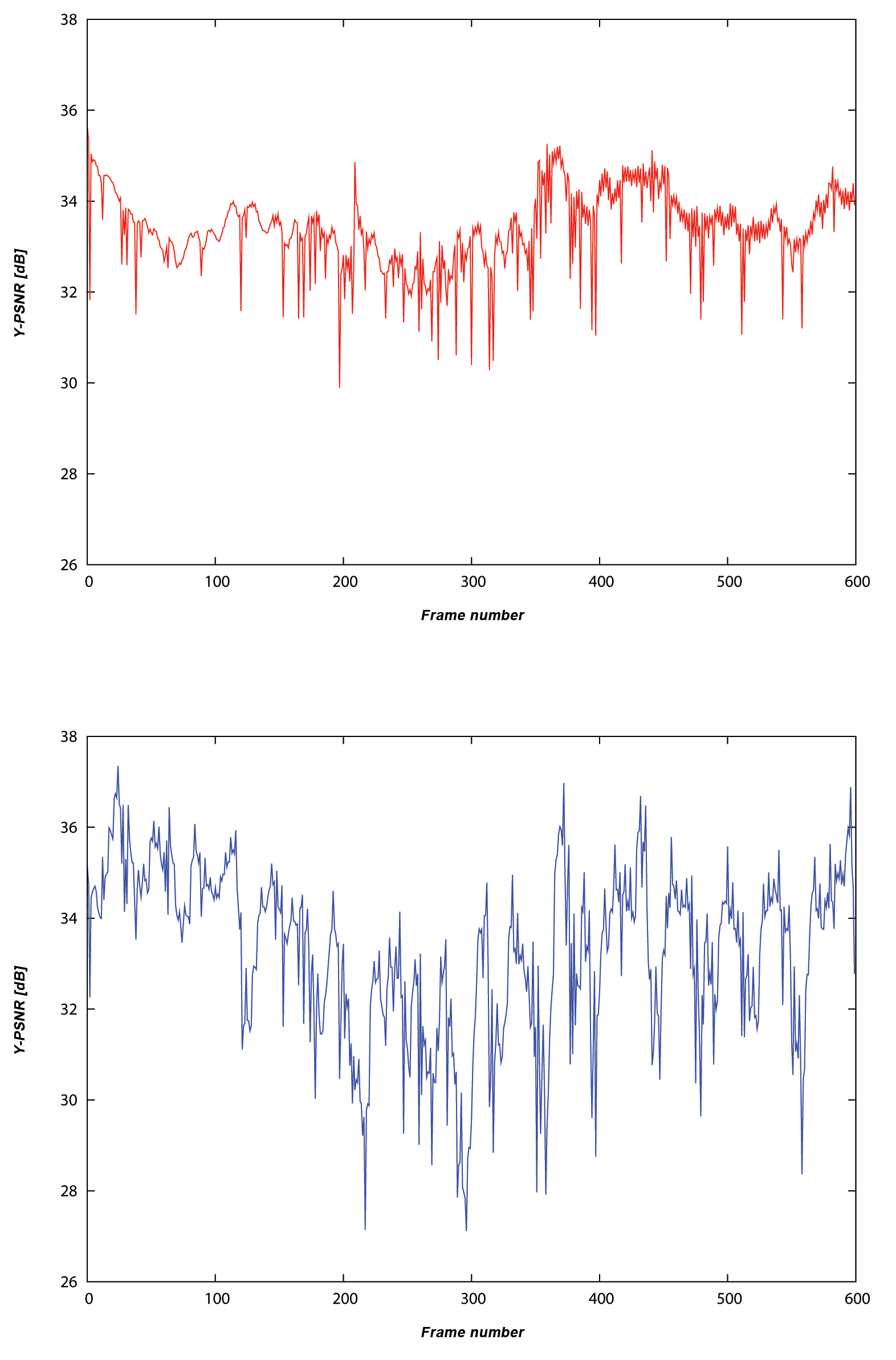}
\caption{Comparison of Y-PSNR of the 'Crew' sequence (\textit{top} x264, \textit{bottom} SVC).}
\label{fig:rc_comp}
\end{figure}

During the quality comparison, a particular flaw in the SVC CODEC became apparent: The rate control. Even though the requested bitrate is delivered in most cases quite accurately, the resulting quality can be unstable under certain conditions.
Figure \ref{fig:rc_comp} shows the Y-PSNR values over the whole 'Crew' sequence for x264 and SVC.

While the maximum fluctuation amplitude of x264 is about $5$ dB, the
SVC CODEC reaches about $10$ dB. Even though the PSNR is inherently
fluctuating due to the hierarchical B-frame structure inside a GOP,
this usually accounts only for about $3$ dB fluctuation. There is
also another significant flaw in SVC rate control: In certain short
sequences, the CODEC tends to distribute too much bitrate at the
beginning of the sequence. This is followed by an excessive increase
of quantization at the end of the file to keep the bitrate inside
the given boundaries. Of course, visual quality suffers
significantly from this non uniform bitrate distribution as short
subjective assessments showed.


It is however noteworthy that this particular behavior did not occur in every sequence, the 'Foreman' and 'City' sequences are, for example, not affected. Further research would be necessary to exactly locate the cause of this problem.

\subsubsection{Encoding speed test}
\label{speed}
The encoding time is measured on two different test systems to evaluate the impact of different CPU speeds and capabilities on SVC encoding. The details of both test systems are listed in tables \ref{tab:test_sys_1} and \ref{tab:test_sys_2}.

\begin{table}[H]
\begin{center}
\begin{tabular}{|l | l|}\hline
OS&Microsoft Windows Vista Business\\
&64-Bit, Version: 6.0.6001 SP1\\\hline
CPU&Intel\textregistered ~Core\texttrademark ~2 Quad Q9450\\
&4$\times$2.66 GHz\\\hline
RAM&4096 MB DDR2 800\\
&Timings: 5-5-5-18\\\hline
BIOS&American Megatrends\\
&Inc. V1.8, 24.01.2008\\\hline
HDD&Samsung Spinpoint T166, 320 GB,\\
&7200 RPM, 16 MB Cache\\\hline
Video Adapter&NVIDIA GeForce 8800 GTS 512\\\hline
\end{tabular}
\caption{Hardware configuration for test system 1.}
\label{tab:test_sys_1}
\end{center}
\end{table}

\begin{table}[H]
\begin{center}
\begin{tabular}{| l | l |}\hline
OS&Microsoft Windows Vista Business\\
&32-Bit, Version: 6.0.6001 SP1\\\hline
CPU&Intel\textregistered ~Core\texttrademark ~2 Duo T5500\\
&2$\times$1.66 GHz @ 1.00 GHz\\\hline
RAM&2048 MB DDR2 667\\
&Timings: 5-5-5-15\\\hline
BIOS&Phoenix Technologies LTD\\
&23YA, 17.04.2007\\\hline
HDD&Hitachi Travelstar 5K100, 100 GB,\\
&5400 RPM, 8MB Cache\\\hline
Video Adapter&ATI Radeon Xpress 200M\\\hline
\end{tabular}
\caption{Hardware configuration for test system 2.}
\label{tab:test_sys_2}
\end{center}
\end{table}

The following tables show the detailed results for both test systems. Both the absolute times and the relative speedup with System 2 as reference are given.

\begin{table}[H]
\begin{centering}
\begin{tabular}{| c | c | c | c |}\hline
&\multicolumn{3}{c|}{\textbf{CIF}}\\
&Xvid&x264&SVC\\\hline
\textbf{System 1}&1.1&0.9&387.7\\\hline
\textbf{System 2}&4.1&7.2&947.8\\\hline\hline
&\multicolumn{3}{c|}{\textbf{4CIF}}\\
&Xvid&x264&SVC\\\hline
\textbf{System 1}&12.1&7.8&2778.5\\\hline
\textbf{System 2}&42.1&60.8&8155.4\\\hline\hline
&\multicolumn{3}{c|}{\textbf{HD}}\\
&Xvid&x264&SVC\\\hline
\textbf{System 1}&15.3&7.7&3902.1\\\hline
\textbf{System 2}&41.6&60.1&9575.8\\\hline
\end{tabular}
\caption{Average encoding time for CIF, 4CIF and HD resolutions on different computer systems in seconds.}
\label{tab:time_comp}
\end{centering}
\end{table}

As table \ref{tab:time_comp} shows, there are significant differences in speedup between the different CODECs.\\
SVC just seems to profit from the higher core clock of system 1, as the speed scales linearly with the core clock\break
$\left(\frac{1{.}00 GHz}{2{.}66 GHz} = 0{.}376\right)$. Xvid speedup is slightly higher, maybe due to optimizations for the new SSE instruction sets implemented in the quadcore processors. The biggest speed gain can be observed using the x264 CODEC. This is because x264 is the only CODEC that supported multithreaded encoding at the time of testing, so the quadcore processor could be used to its full potential. It has to be mentioned that the new 1.2.1 version of Xvid also supports multithreaded encoding, so the speedup can be expected be on par with x264.

\section{Current SVC flaws}
\label{sec:conclusion}
\subsection{Improvement of existing features}
While the new MPEG-4 SVC CODEC adds many useful features to its predecessor MPEG-4 AVC, some flaws could still be observed during the subjective as well as the objective evaluations. These are described in the next section.

\subsubsection{More reasonable default configuration}
Some parameters of the SVC configuration files are by default not reasonably adjusted. The most important is the value of 'BaseLayerMode', whose default value is '3', which is not even a defined setting as the only possible choices are '0' (= 'AVC compatible base layer with larger DPB size,'), '1' (= 'AVC compatible base layer') or '2' (= 'AVC compatible base layer with sub-sequence SEI messages for supporting temporal scalability'). As a result, it is proposed to change the currently undefined value of '3' to a allowed one. The value '2' is used during all the assessments in this work, as it is the most advanced of all.\\
Although being allowed and defined, the value of '1' for the setting 'GOPSize' is also not reasonable, as it heavily cripples the amount of temporal scalability possible. To understand this, it is necessary to explain the changes in coding structure that come along with the size of the GOP:\\

\begin{table}[H]
\begin{center}
\begin{tabular}{| c | c |}\hline
\textbf{GOPSize}&\textbf{Coding structure}\\\hline
\textbf{1}&\texttt{I~P~P~P~P~P~P~P~P~P~P~P~P~P~P~P~P~P...}\\\hline
\textbf{2}&\texttt{IB~PB~PB~PB~PB~PB~PB~PB~PB...}\\\hline
\textbf{4}&\texttt{IBBB~PBBB~PBBB~PBBB~PB...}\\\hline
\textbf{8}&\texttt{IBBBBBBB~PBBBBBBB~PB...}\\\hline
\textbf{16}&\texttt{IBBBBBBBBBBBBBBB~PB...}\\\hline
\end{tabular}
\caption{Coding structure for different GOP sizes.}
\label{tab:GOP}
\end{center}
\end{table}

As the table above shows, the default 'GOPSize' value of '1' causes that no B-frames at all are used, which leads, on the positive side, as the previous tests have shown, to a shorter encoding time but also to only one temporal level without any scaled substreams. Hence, a change of the default parameter to a value of '8' or '16' is purposed.\\
Because the encoding speed of SVC is currently low, the default parameter '0' (= 'BlockSearch') of 'SearchMode' is also not considered to be reasonable, as the quality gain of the blocksearch algorithm is - compared to the severely higher encoding time needed - only marginal. In order to significantly improve the encoding speed the default value of 'SearchMode' should be switched to '4' (= 'FastSearch') .

\subsubsection{Improve encoding speed}
The previous test have shown that the current MPEG-4 SVC version has a much lower encoding speed than the other tested CODECs. Firstly, it needs to be mentioned again that this is to be expected, as SVC is still in development status, but two main reasons can be identified and are explained in the following.

\paragraph{Multithreading}
The benefit of multithreading support becomes more and more visible in modern computer systems, because multicore configurations are already commonly found in private environments today. If a similar encoding speed gain as in x264 when using multithreading is proclaimed, the encoding speed would approximately be accelerated linearly with the number of available CPUs.\\
Although this increase would still not be sufficient to keep up with the other CODECs, it would obviously be a step in the right direction.\\
The main challenge in this process would be a reasonable parallelization of the encoding steps to correctly and effectively split the work among the available CPUs, which would especially concern the motion estimation process, as the evaluation has shown.

\paragraph{Performance improvements of the motion estimation}
To further decrease the encoding time needs, it would be essential to optimize the performance of the motion estimation algorithms. As already noted in \cite{bb15340}, the currently employed motion estimation technique achieves the best quality possible. However, the computation complexity is very high, which obstructs it from practical use. \cite{bb15340} also proposes a fast mode decision algorithm for inter-frame coding as a solution, which relies on the mode-distribution correlation between the base and enhancement layers. Using this algorithm, an average encoding time reduction of 53\% can be achieved, while the visual quality and bitrate only suffer minor impacts.\\
The encoding speed as well as the usability would highly benefit if the proposed or similar motion estimation algorithms would be included in the SVC CODEC.

\subsubsection{Enhanced, stable rate control mechanism}
As shown in the synthesis of the CODEC comparison evaluation, the SVC rate control feature still has minor flaws, which manifest in two ways:\\
Firstly, the sequences encoded using rate control have a much more unstable PSNR value resulting in quality fluctuations.\\
Secondly, some sequences show severe quality degradation at the last frames, which is supposably also caused due to the inability of rate control to correctly adjust the bitrate throughout the whole sequence.\\

\begin{figure}[H]
\centering
\includegraphics[width=0.35\textwidth]{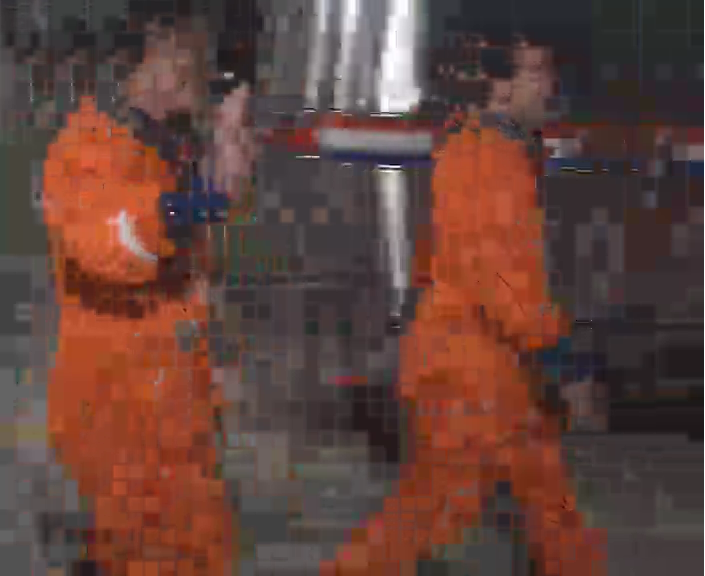}
\caption{Rate control introduced blocking artifacts at the end of a sequence.}
\label{fig:rate_end}
\end{figure}

Because the exact reasons for these behaviors could not be precisely pinpointed in the tests, no concrete proposal for improvement can be given here. Still, improvements in this area are regarded as necessary.

\subsection{Additional useful features}
In the next section, additional features, that are not implemented in the current SVC release, but would be useful, are described.

\subsubsection{Variable, content-dependent framerate}
As scalable video technology is especially advantageous in streaming media environments, a useful new technique, which is already used by other video CODECs, would be the usage of a content-aware dynamic framerate.\\
An example for the successful implementation of variable framerate is the Blackbird CODEC used in the FORscene system developed by Forbidden Technologies plc., which is optimized for video transmission over heterogeneous networks. Because the CODEC is fully proprietary, no further information can be given here.\\
The basic idea of variable content-dependent framerate is that a reduced temporal level does not impair scenes with no or very low movement, which was already proven by~\cite{temp_metric}. There could be two main positive results when reducing the framerate: Either the file size of the video sequence could be reduced, or - if the size remains constant - the SNR quality would benefit respectively.

\subsubsection{2-Pass encoding mode}
2-pass encoding strategies have been implemented in most modern CODECs, for example Xvid or x264 which have been examined earlier. 2-pass encoding works by first analyzing the videos complexity (first pass), after that the available bitrate is distributed dynamically to achieve maximum quality (second pass). This is especially useful for archiving purposes, as high bitrate 'spikes' are not of concern. In contrast, when using 2-pass mode for streaming applications, special care has to be taken not to overload the connection. This can be done on client side or while encoding the video. On client side, high differences in video bitrate can be compensated by using large buffers, of course this also has downsides: First, filling these buffers can take a certain amount of time, so the user has to wait before the requested video starts. Second, the memory required for storing a high amount of video frames is not always available, especially in highly mobile devices. If the problem of bitrate spikes is addressed while encoding, a threshold value has to be defined as an absolute maximum bitrate, so that the bandwidth of the connection is always capable of delivering the video stream.\\
Implementing this feature into SVC would primarily benefit its suitability for archiving storage. Of course, the poor rate control of SVC would also benefit from the bitrate distribution algorithms in 2-pass mode. In spite of this fact, it is essential that single pass rate control of SVC is improved, as 2-pass encoding mode is not suited for realtime encoding.

\subsection{Conclusion}
The extensive tests conducted in this work showed that the new scalable video coding extensions provide significant improvement in terms of adaptability of the video stream. This is especially important in the modern, heterogeneous network conditions caused by the growing number of mobile multimedia devices. In contrast, there is also a growing demand for high quality digital video, mostly for the emerging high definition television standard. Using the scalability features of SVC, both of these demands can be met simultaneously, while at the same time saving bitrate compared to the storage of separate videos tailored for each device. Further, using the combination of media aware network components and IP multicasting could provide an enormous potential for saving upstream bandwidth for video servers.\\
However, there are also several features that still need improvement. First and foremost, the encoding speed of the SVC reference encoder is far too slow. Two methods to speed up the encoding are already proposed before. The successful acceleration of the encoding process is by far the most pressing matter, as usage at current speed levels is not feasible in large scale. Additionally, several optimizations and other new useful features are proposed in the previous section.\\
Concluding, SVC is a promising new extension to the MPEG CODEC family. If the most severe issues are addressed, it is likely to significantly improve the viewing experience of digital video consumers.

\nocite{SUB-MPEG4}
\nocite{bb9878}
\nocite{COMP-VID}
\nocite{MSU-SUB-MET}

\nocite{Voran91}
\nocite{bb14503}
\nocite{GIROD_ERROR}
\nocite{bb14503}
%
\bibliographystyle{abbrv}
\bibliography{bib}  

\begin{thebibliography}{10}

\bibitem{bb14503}
E.~Akyol, A.~M. Tekalp, and M.~R. Civanlar.
\newblock [motion-compensated temporal filtering within the h.264/avc standard.

\bibitem{SUB-MPEG4}
T.~Alpert, V.~Baroncini, D.~Choi, L.~Contin, R.~Koenen, F.~Pereira, and
  H.~Peterson.
\newblock {Subjective evaluation of MPEG-4 video codec proposals:
  Methodological approach and test procedures}.
\newblock {\em SP:IC}, 9(4):305--325, May 1997.

\bibitem{FRACT}
M.~Bach.
\newblock {Freiburg Visual Acuity, Contrast \& Vernier Test ('FrACT')}.
\newblock \url{http://www.michaelbach.de/fract/index.html}, 2002.

\bibitem{bb9878}
M.~A.~J. Barzilay, J.~R. Taal, and R.~L. Lagendijk.
\newblock {Subjective Quality Analysis of Bit Rate Exchange Between Temporal
  and SNR Scalability in the MPEG4 SVC Extension}.
\newblock In {\em International Conference on Image Processing}, pages II:
  285--288, 2007.

\bibitem{COMP-VID}
J.~Biström.
\newblock {Comparing Video Codec Evaluation Methods for Handheld Digital TV}.
\newblock In {\em {T-111.590 Research Seminar on Digital Media}}, 2005.

\bibitem{msu_psnr}
{CS MSU Graphics \& Media Lab Video Group}.
\newblock {MSU Quality Measurement Tool: FAQ}.
\newblock \url{http://compression.ru/video/quality_measure/vqmt_faq_en.html#9}.

\bibitem{RTP1}
A.~Durresi and R.~Jain.
\newblock {RTP, RTCP, and RTSP - Internet Protocols for Real-Time Multimedia
  Communication}.
\newblock In R.~Zurawski, editor, {\em The Industrial Information Technology
  Handbook}, pages 1--11. CRC Press, 2005.

\bibitem{temp_metric}
R.~Feghali, D.~Wang, F.~Speranza, and A.~Vincent.
\newblock {Quality Metric for Video Sequences with Temporal Scalability}.
\newblock In {\em International Conference on Image Processing}, pages III:
  137--140, 2005.

\bibitem{ISO_VERI}
I.~O. for Standardisation.
\newblock Svc verification test report. iso/iec jtc 1/sc 29/wg 11 n9577, 2007.

\bibitem{GIROD_ERROR}
B.~Girod.
\newblock {What's wrong with Mean-Squared Error}.
\newblock {\em SP:IC}, 15:207--220, 1993.

\bibitem{INST-RUND}
{Institut für Rundfunktechnik}.
\newblock {ITU-R BT.500 Recommendation and SAMVIQ (ITU-R BT.700)}.
\newblock \url{http://radio.irt.de/docs/subjectiveVQA.html}, 2005.

\bibitem{ITU-BT500}
{International Telecommunication Union}.
\newblock {\em {ITU-R Recommendation BT.500 - Methodology for the Subjective
  Assessment of the Quality of Television Pictures}}.
\newblock {International Telecommunication Union}, {Geneva}, 2004.

\bibitem{color}
{Jeff Rabin (Visual Function Laboratory Ophthalmology Branch / USAF School of
  Aerospace Medicine)}.
\newblock {Color Vision Fundamentals}.
\newblock
  \url{http://colorvisiontesting.com/RABIN%20slide%20presentation%20for%20webp%
age.ppt}, 1998.

\bibitem{softman}
{Joint Video Team (ISO/IEC Moving Pictures Experts Group (MPEG) and ITU-T Video
  Coding Experts Group (VCEG))}.
\newblock {\em JSVM Software Manual}, 2008.

\bibitem{SAMVIQ-EBU}
F.~Kozamernik, V.~Steinman, P.~Sunna, and E.~Wyckens.
\newblock {\em {SAMVIQ - A New EBU Methodology for Video Quality Evaluations in
  Multimedia}}.
\newblock Amsterdam, 2004.

\bibitem{bb15340}
H.~Li, Z.~G. Li, and C.~Wen.
\newblock {Fast Mode Decision Algorithm for Inter-Frame Coding in Fully
  Scalable Video Coding}.
\newblock {\em {IEEE Trans. Circuits and Systems for Video Technology}},
  16(7):889--895, July 2006.

\bibitem{RTP2}
C.~Liu.
\newblock {Multimedia Over IP: RSVP, RTP, RTCP, RTSP}.
\newblock \url{http://vodka.lfcia.org/docs/RTP/ip_multimedia.pdf}, Mar.~19
  2001.

\bibitem{vqeg_results}
A.~M. Rohaly, P.~J. Corriveau, J.~M. Libert, A.~A. Webster, V.~Baroncini,
  J.~Beerends, J.-L. Blin, L.~Contin, T.~Hamada, D.~Harrison, A.~P. Hekstra,
  J.~Lubin, Y.~Nishida, R.~Nishihara, J.~C. Pearson, A.~F. Pessoa, N.~Pickford,
  A.~Schertz, M.~Visca, A.~B. Watson, and S.~Winkler.
\newblock {Video Quality Experts Group: current results and future directions}.
\newblock In K.~N. Ngan, T.~Sikora, and M.-T. Sun, editors, {\em Visual
  Communications and Image Processing 2000}, volume 4067 of {\em Proceedings of
  SPIE}, pages 742--753. SPIE, 2000.

\bibitem{soft_manual}
J.~V. Team.
\newblock {\em SVC Software Manual}, 2008.

\bibitem{comparison2007}
D.~Vatolin, D.~Kulikov, and A.~Parshin.
\newblock {Video MPEG-4 AVC/H.264 Codecs Comparison}.
\newblock Dec. 2007.

\bibitem{comparison2006}
D.~Vatolin, A.~Parshin, O.~Petrov, and A.~Titarenko.
\newblock {MSU Subjective Comparison of Modern Video Codecs}.
\newblock Jan. 2006.

\bibitem{MSU-SUB-MET}
D.~Vatolin, M.~Smirnov, A.~Ratushnyak, V.~Yoockin, and A.~Andreev.
\newblock {MSU Perceptual Video Quality tool - Description of methods of
  subjective video quality evaluation}.
\newblock 2008.

\bibitem{Voran91}
S.~D. Voran.
\newblock {The Development of Objective Video Quality Measures that Emulate
  Human Perception}.
\newblock {\em IEEE Globecom 91}, 3 volumes:1776--1781, Dec. 1991.

\bibitem{ssim2004}
Z.~Wang, A.~C. Bovik, H.~R. Sheikh, and E.~P. Simoncelli.
\newblock {Image Quality Assessment: From Error Visibility to Structural
  Similarity}.
\newblock {\em IEEE Trans. Image Processing}, 13(4):600--612, Apr. 2004.

\bibitem{PLUGE}
{Wavelength Media}.
\newblock {\em {Pluge Test Pattern}}.
\newblock \url{http://www.mediacollege.com/video/test-patterns/pluge.html}.

\bibitem{bb15205}
M.~Wien, H.~Schwarz, and T.~Oelbaum.
\newblock {Performance Analysis of SVC}.
\newblock {\em IEEE Trans. Circuits and Systems for Video Technology},
  17(9):1194--1203, Sept. 2007.

\end{thebibliography}
%
%
\end{document}